\documentclass[journal]{IEEEtran}

\usepackage{epsf,psfrag,amssymb,amsfonts,color,cite,fancybox}
\usepackage[mathscr]{eucal}
\usepackage[dvips]{graphicx}
\usepackage[dvipsnames]{xcolor}
\usepackage{caption}
\usepackage{subcaption}
\usepackage{ifpdf}
\usepackage{longtable}	
\usepackage{multicol}
\usepackage{float}
\usepackage{epstopdf}
\usepackage{bm}
\usepackage{multirow}
\usepackage[scientific-notation=true]{siunitx}
\usepackage{url}

\usepackage{algorithm}
\usepackage{algpseudocode}

\ifCLASSINFOpdf

\else

\fi

\usepackage[cmex10]{amsmath}
\begin{document}

\bstctlcite{IEEEbib:BSTcontrol}

\title{Online Measurement-Based Estimation of Dynamic System State Matrix in Ambient Conditions}

\author{Hao Sheng,~\IEEEmembership{Member,~IEEE},
Xiaozhe Wang,~\IEEEmembership{Member,~IEEE}

\thanks{This work is supported by Fonds de recherche du Qu\'{e}bec--Nature et technologies (FRQNT) New university researchers start up program and Natural Sciences and Engineering Research Council (NSERC) Discovery Grant RGPIN-2016-04570.}
\thanks{Hao Sheng and Xiaozhe Wang are with the Department of Electrical and Computer Engineering, McGill University, Montr\'{e}al, QC H3A 0G4, Canada. emails: shenghao@tju.edu.cn, xiaozhe.wang2@mcgill.ca.}
}

\markboth{IEEE Transactions on Smart Grid, Preprint, May 2019}%
{Sheng \MakeLowercase{\textit{et al.}}:
Online Measurement-Based Estimation of Dynamic System State Matrix in Ambient Conditions
}

\maketitle


%
\begin{abstract}
In this paper, a purely measurement-based method is proposed to estimate the dynamic system state matrix by applying the regression theorem of the multivariate Ornstein-Uhlenbeck process. The proposed method employs a recursive algorithm to minimize the required computational effort, making it applicable to the real-time environment. One main advantage of the proposed method is model independence, i.e., it is independent of the network model and the dynamic model of generators. Among various applications of the estimated matrix, detecting and locating unexpected network topology change is illustrated in details. Simulation studies have shown that the proposed measurement-based method can provide an accurate and efficient estimation of the dynamic system state matrix under the occurrence of unexpected topology change.
Besides, various implementation conditions are tested to show that the proposed method can provide accurate approximation despite measurement noise, missing PMUs, and the implementation of higher-order generator models with control devices.
\end{abstract}


\IEEEpeerreviewmaketitle


%
\section{Introduction}
\IEEEPARstart{T}{he} primary goal of the power system operators is to maintain the system in the secure, reliable and efficient operating state despite varying operating conditions. Achieving this goal necessitates continuous monitoring of the system condition, conducting the periodic security analysis, and determining necessary control actions. The power flow Jacobian matrix and the dynamic system state matrix (also termed as the system dynamic state Jacobian in \cite{Pai:1990}) play essential roles in power system analysis since they contain invaluable information regarding the existence of the steady-state equilibrium point and the stability of the power system dynamic model, respectively. The critical difference between these two matrices is that the dynamic system state matrix entails the dynamic information such as the stability \cite{Pai:1990}, the oscillation modes \cite{PKundur94a}, the coherency groups \cite{JHChow13a} of the dynamical power system, which cannot be achieved from the power flow Jacobian matrix for general cases \cite{Pai:1990}. 
Note that these matrices can be easily obtained if the power system model and topology are known and accurate. However, the parameters of components may be inaccurate due to the errors introduced in the modeling stage (e.g., inaccurate manufacturing data and line length) or maintenance stage. 
Moreover, the topology may be corrupted due to erroneous telemetry from a remotely monitored circuit breaker. The above factors may lead to inaccurate or even incorrect results of static and dynamic security analysis. 

Broad adoption of phasor measurement units (PMUs) and wide-area measurement system (WAMS) technology \cite{JDLRee10a} provides a great opportunity to bypass this issue and gives birth to a new class of methods termed as measurement-based methods for power system monitoring, analysis, and control. 
The unique feature of these methods is that they rely mainly on high-speed, time-synchronized PMU data and are partially or entirely model-free. Significant attention have been giving to the PMU-based applications such as robust state estimation (e.g., \cite{JZhao16a}), generator dynamic state estimation (e.g., \cite{NZhou15a}), fault detection and location (e.g., \cite{CWLiu12a}, \cite{MMajidi17a}), inter-area oscillations (e.g., \cite{JMa10a}, \cite{SNabavi15a}), dynamic equivalent (e.g., \cite{Vittal:2012}, \cite{AChakrabortty11a}), model calibration (e.g.,\cite{AAHajnoroozi15a}), etc. 

The broad adoption of PMUs also offers an alternative way to estimate the matrices mentioned above solely from measurement data. In previous literature, authors of \cite{Chen:2016} utilized high-speed synchronized voltage and current phasor data to determine the power flow Jacobian matrix through linear total least-square estimation; authors of \cite{PLi18a} exploited the sparsity of the power flow Jacobian matrix and transformed its estimation into a sparse-recovery problem via compressed sensing; authors of \cite{Demarco:2016} estimated the largest singular value of the inverse Jacobian matrix utilizing PMU data, which serves as a voltage stability indicator. However, the matrix considered in these works is the power flow Jacobian matrix where no dynamic information of the system is embedded. 
In our previous work\cite{Wangxz:2017TRWS}, \cite{Wangxz:2017ISCAS}, we have proposed a hybrid measurement and model-based method to estimate the dynamic system state matrix which carries significant information about the dynamics of the system. The proposed approach is not purely measurement-based as we assume the inertia constants $M$ are known, but no information of network topology and parameters is needed.

In this paper, we make one further step by proposing a purely measurement-based method for estimating the dynamic system state matrix. Compared to \cite{Wangxz:2017TRWS},\cite{Wangxz:2017ISCAS}, the proposed approach is entirely model-free, i.e., it does not require any information about the network model and the dynamic generator model. This assumption relief is significant as the inertia and damping constants are also subject to inaccuracy and unavailability.
To remove this assumption, the regression theorem of multivariate Ornstein-Uhlenbeck process \cite{Gardiner:2009} is utilized to estimate the dynamic state Jacobian matrix from the stationary time correlation and covariance matrix of measurements. To enable online application, a recursive estimation algorithm is developed to determine the matrix in near real-time. 
In sum, the contributions of the paper are presented as below:
\begin{itemize}
\item A model-free, purely measurement-based method is proposed to estimate the dynamic system state matrix by applying the regression theorem of the multivariate Ornstein-Uhlenbeck process. 
\item A fast recursive estimation algorithm is developed to minimize the required computational effort, enabling the application in the near real-time environment.
\item An accurate estimate of the dynamic system state matrix can be obtained despite measurement noise, missing PMUs and the implementation of higher-order generator models with control devices. The topology change can be identified and located to the closest generator bus in terms of electrical distance.
\end{itemize}

The rest of the paper is organized as follows. Section \ref{sectionmodel} introduces the stochastic power system dynamic model and elaborates the proposed recursive measurement-based method for dynamic system state matrix. Section \ref{casestudy} presents the simulation studies to illustrate the feasibility and accuracy of the proposed method under various implementation conditions and demonstrate the application of the estimated matrix in detecting and locating network topology change.


%
\section{The system model and the proposed methodology}\label{sectionmodel}

The general power system dynamic model can be described as below:
\small
\begin{eqnarray}
\dot{\bm{x}}&=&\bm{f}({\bm{x},\bm{y}})\label{fast ode}\\
\bm{0}&=&\bm{g}({\bm{x},\bm{y}})\label{algebraic eqn}
\end{eqnarray}
\normalsize
where (\ref{fast ode}) describes dynamics of generators, excitation systems, load dynamics, etc.,
and (\ref{algebraic eqn}) describes the transmission system and the static behaviors of passive devices. $f$ and $g$ are continuous functions; vectors $\bm{x}\in\mathbb{R}^{n_{\bm{x}}}$ and $\bm{y}\in\mathbb{R}^{n_{\bm{y}}}$ are the corresponding state variables (e.g., generator rotor angles and speeds) and algebraic variables (e.g., terminal bus voltage magnitudes and angles) \cite{Wangxz:CAS}.

In this paper, we are interested in small signal stability when the power system operates around its steady state, i.e., in ambient conditions. Under this condition, the dynamics of generators or aggregated generators can be represented by the swing equations. 
Thus (\ref{fast ode})-(\ref{algebraic eqn}) take the following form \cite{Bialek:book}: 
\small
\begin{eqnarray}
\dot{\bm{\delta}}&=&\bm{\omega}\label{swing-1}\\
M\dot{\bm{\omega}}&=&\bm{P_m}-\bm{P_e}-{D}\bm{\omega}\label{swing-2}
\end{eqnarray}
\normalsize
\noindent with
\small
\begin{equation}
P_{e_i}=\sum_{j=1}^{n}E_iE_j\left[G_{ij}\cos({\delta}_i-{\delta}_j)+B_{ij}\sin(\delta_i-{\delta}_j)\right]\label{swing-3}
\end{equation}
\normalsize
\noindent where the notations are as in Table \ref{notation}.
\begin{table}[ht]\normalsize
\begin{tabular}{ll}
  $\delta_i$ & generator rotor angle\\
  $\omega_i$ & generator angular frequency\\
  $M_i$ & inertial constant \\
  $P_{mi}$ & mechanical power input \\
  $P_{ei}$ & real power injection \\
  $D_i$ & damping coefficient \\
  $n$ & total number of buses\\
  $E_i$ & the voltage magnitude behind the transient reactance\\
  $G_{ij}$ & equivalent conductance between bus $i$ and $j$\\
  $B_{ij}$ & equivalent susceptance between bus $i$ and $j$ 
  \end{tabular}
 \caption{Notations for the power system dynamic model}\label{notation}
\end{table}

In this paper, we make the common assumption \cite{Pal:2010} that the active powers of loads are perturbed by independent Gaussian variation from their base loadings. 
As shown in \cite{Crow:2013},\cite{Nwankpa:2000}, the load variations can be described by the stochastic perturbations at the diagonal elements of the reduced admittance matrix by $Y(i,i)=G_{ii}+jB_{ii}=Y_{ii}(1+\sigma_i {\bm{\xi}})\angle{\phi_{ii}}$, where $\bm{\xi}=[\dot{W_1},...,\dot{W_n}]^T$ is a vector of independent standard Gaussian random variable, and $\sigma_i^2$ denotes the variance of stochastic variation. Similar to the approach in \cite{Crow:2013}, we assume that the power factor ({and subsequently $\phi_{ii}$}) remains unchanged, i.e., the load variations only reveal in the magnitude of admittance and not in the angle such that the nonlinearity does not kick in. Therefore, the stochastic power system model considering stochastic load variations can be described by:
\small
\begin{eqnarray}
\dot{\bm{\delta}}&=&\bm{\omega}\label{swingrandom-1}\\
M\dot{\bm{\omega}}&=&\bm{P_m}-\bm{P_e}-{D}\bm{\omega}-E^2G\Sigma\bm{\xi}\label{swingrandom-2}
\end{eqnarray}
\normalsize
where $P_{e_i}$ is given in (\ref{swing-3}), $E=\mbox{diag}([E_{1},...,E_{n}])$, $G=\mbox{diag}([G_{11},...,G_{nn}])$, $\Sigma=\mbox{diag}([\sigma_1,...,\sigma_n])$. 
Note that in the above formulation, system loads are modeled as constant impedances, since we assume that the system dynamics are dominated by generator angles. More realistic load models such as ZIP loads may be incorporated by adding additional terms in (\ref{swing-3})\cite{Elahi:1983}, which on the other hand, does not affect the overall formulation.

Linearizing (\ref{swingrandom-1})-(\ref{swingrandom-2}) around the steady state, we obtain:
\small
\begin{equation}
\label{swing-matrix}
\begin{aligned}
\left[\begin{array}{c}{\Delta}\dot{\bm{\delta}}\\{\Delta}\dot{\bm{\omega}}\end{array}\right]&=
\left[\begin{array}{cc}{{0}}&{I_n}\\-M^{-1}\frac{\partial{\bm{P_e}}}{\partial{\bm{\delta}}}&-M^{-1}D\end{array}\right]
\left[\begin{array}{c}{{\Delta}\bm{\delta}}\\{{\Delta}\bm{\omega}}\end{array}\right] \\
&+\left[\begin{array}{c}0\\-M^{-1}E^2G\Sigma\end{array}\right]\bm{\xi}
\end{aligned}
\end{equation}
\normalsize
where $I_n$ is an identity matrix of size $n$. If we set
\small
\begin{equation}
\begin{gathered}
\bm{x}=\left[\begin{array}{c}{{\Delta}\bm{\delta}} \\ {{\Delta}\bm{\omega}}\end{array}\right],\enskip
A=\left[\begin{array}{cc}{{0}}&{I_n}\\-M^{-1}\frac{\partial{\bm{P_e}}}{\partial{\bm{\delta}}}&-M^{-1}D\end{array}\right], \\
B=\left[\begin{array}{c}{0}\\{-M^{-1}E^2G\Sigma}\end{array}\right]
\end{gathered}
\end{equation}
\normalsize
then (\ref{swing-matrix}) can be represented by the following set of stochastic differential equations which is indeed a vector Ornstein-Uhlenbeck process \cite{Gardiner:2009}:
\small
\begin{equation}
\label{eq:OU}
\dot{\bm{x}}=A\bm{x}+B\bm{\xi}
\end{equation}
\normalsize

The stationary covariance matrix is defined as:
\small
\begin{equation}
C=\left\langle\left[\bm{x}(t)-\mu_{\bm{x}}\right]\left[\bm{x}(t)-\mu_{\bm{x}}\right]^T\right\rangle=\left[\begin{array}{cc}C_{\bm{\delta}{\bm{\delta}}}&C_{\bm{\delta}{\bm{\omega}}}\\C_{\bm{\omega}{\bm{\delta}}}&C_{\bm{\omega}{\bm{\omega}}}\end{array}\right] \end{equation}
\normalsize
where $\mu_{\bm{x}}$ denotes the mean; the $\tau$-lag time correlation matrix is defined as:
\small
\begin{equation}
G(\tau)=\left\langle\left[\bm{x}(t+\tau)-\mu_{\bm{x}}\right]\left[\bm{x}(t)-\mu_{\bm{x}}\right]^T\right\rangle
\end{equation}
\normalsize

The regression theorem of multivariate Ornstein-Uhlenbeck process states that the following relation holds \cite{Gardiner:2009}: 
\small
\begin{equation}
\frac{d}{d\tau}\left[G(\tau)\right]=-AG(\tau)
\end{equation}
\normalsize
if the dynamic system state matrix $A$ is stable. In other words, if the system is in ambient conditions such that $A$ is stable, the dynamic system state matrix can be solved by:
\small
\begin{equation}
\label{systemmatrix}
A=\frac{1}{\tau}\log\left[G(\tau)C^{-1}\right]
\end{equation}
\normalsize
Equation (\ref{systemmatrix}) indeed is the theoretical basis of the proposed measurement-based method, which links the statistical properties of the measurements tactfully up with the physical model knowledge. Note that estimating $A$ via (\ref{systemmatrix}) is purely measurement-based since no knowledge of dynamic model parameters (e.g., damping and inertia constants) and network model (e.g., topology and line parameters) is assumed. Particularly, the proposed method requires no information of inertia constants compared to the hybrid method in \cite{Wangxz:2017TRWS}. It is also worth mentioning that the proposed method is different from system identification methods in the sense that the estimated matrix $A$ corresponds to the true dynamic system state matrix of the physical model whereas the matrix $A$ acquired from typical system identification methods does not necessarily correspond to the true dynamic system state matrix. 

In Section \ref{recursive method}, we will develop a fast recursive estimation method based on (\ref{systemmatrix}) to estimate the dynamic system state matrix in near real-time. Before that, we will discuss the estimation of the time correlation matrix $G(\tau)$ and the covariance matrix $C$ from limited data set.

\subsection{Estimation of $G(\tau)$ and $C$}

Ideally, $C$ and $G$ are calculated from an infinite number of data which is impossible in practice. Hence, the sample mean $\bar{\bm{x}}$, sample covariance matrix $\hat{C}$ and sample correlation matrix $\hat{G}$ calculated from finite data set is typically used as follows:
\small
\begin{equation}
\label{samplemean}
\bar{\bm{x}}=\frac{1}{N}\sum_{i=1}^{N}\bm{x}_i
\end{equation}
\normalsize
\small
\begin{equation}
\label{samplecovariance}
\hat{C}=\frac{1}{N}\left(F-\bar{\bm{x}}\boldsymbol{1}^T_N\right)\left(F-\bar{\bm{x}}\boldsymbol{1}^T_N\right)^T
\end{equation}
\normalsize
\small
\begin{equation}
\label{samplecorrelatoin}
\hat{G}(\triangle t)=\frac{1}{N}(F_{2:N}-\bar{\bm{x}}\boldsymbol{1}^T_{N-1})(F_{1:N-1}-\bar{\bm{x}}\boldsymbol{1}^T_{N-1})^T 
\end{equation}
\normalsize
where $N$ is the sample size, $F=[\bm{x}_1, \bm{x}_2,\cdots,\bm{x}_N]$ is a $M_s \times N$ matrix assuming there are $M_s$ state variables, $F_{i:j}$ denotes the submatrix of $F$ from $i$ to $j$ columns, $\boldsymbol{1}_N$ is an $N$ by 1 vector of ones, and $\triangle t$ is the sampling time step. Finally, the system state matrix can be estimated from:
\small
\begin{equation}
\label{samplesystemmatrix}
A=\frac{1}{\triangle t}\log\left[\hat{G}(\triangle t)\hat{C}^{-1}\right]
\end{equation}
\normalsize

\subsection{The Recursive Estimation Algorithm}\label{recursive method}
In order to enable online estimation, the recursive method is preferred which does not require calculating the basic equations over and over again. Once the initial $\hat{C}$ and $\hat{G}$ are estimated from a pre-defined length of data, they can be updated recursively using new data. For instance, the sample mean satisfies:
\small
\begin{equation}
\begin{aligned}
\bar{\bm{x}}_N&=\frac{1}{N}\sum_{i=1}^N\bm{x}_i=\frac{1}{N}\left(\sum_{i=1}^{N-1}\bm{x}_i+\bm{x}_N\right) \\
&=\left(1-\frac{1}{N}\right)\bar{\bm{x}}_{N-1}+\frac{1}{N}\bm{x}_N
\end{aligned}
\end{equation}
\normalsize
Similar recursion for $\hat{C}$ and $\hat{G}$ can also be obtained.

Typically, we may generalize the recursion formulas by replacing $\frac{1}{N}$ by $\alpha$. How to choose the smoothing factor $\alpha$ will be discussed in the subsequent section. Moreover, the inverse covariance matrix can be calculated without inverting directly using Sherman-Morrison formula \cite{GHGolub12a}:
\small
\begin{equation}
\begin{aligned}
\hat{C}_N^{-1}&=\left[(1-\alpha)\hat{C}_{N-1}+\alpha\bm{z}_N\bm{z}_N^T\right]^{-1} \\
&=\frac{1}{(1-\alpha)}\left[\hat{C}_{N-1}^{-1}-\alpha\frac{\hat{C}_{N-1}^{-1}\bm{z}_N\bm{z}_N^T\hat{C}_{N-1}^{-1}}{1+\alpha\bm{z}_N^T\hat{C}_{N-1}^{-1}\bm{z}_N}\right]
\end{aligned}
\end{equation}
\normalsize

Before elucidating the algorithm, we make the following assumptions. First, PMUs are installed at all power plants where generators are connected, which is a reasonable assumption according to the basic guidelines for placing PMUs\cite{Chow:2011}. Second, we can use the PMUs to estimate the values of rotor angle $\bm{\delta}$ and rotor speed $\bm{\omega}$ in ambient conditions as explained in \cite{Zhou:2011},\cite{Angel:2003},\cite{Liu:2011}. Note that such task may be difficult in transient condition and thus require dynamic system state estimation techniques, yet is achievable in ambient conditions \cite{Zhou:2011}. 
Based on these assumptions, the following recursive algorithm is able to estimate the dynamic system state matrix in near real-time.

\noindent\textbf{Step 1.} Given a pre-defined window length $N$ data, initialize $\bar{\bm{x}}_0$, $\hat{C}_0$ and $\hat{G}_{0}(\triangle t)$ for $j=0$ by using (\ref{samplemean})-(\ref{samplecorrelatoin}), compute the inverse matrix $\hat{C}_0^{-1}$ and calculate the initial estimation for the dynamic system state matrix $A_0$:
\small
\begin{equation}
A_0=\frac{1}{\triangle t}\log\left[\hat{G}_{0}(\triangle t)\hat{C}_0^{-1}\right]\label{eq:algorithm-1}
\end{equation}
\normalsize

\noindent\textbf{Step 2.} For $j=1,2,\cdots$, perform the following:
\begin{itemize}
\item a) Update the sample mean, the sample correlation matrix and the inverse of the sample covariance matrix:
\small
\begin{equation}
\label{eq:recurformula}
\begin{gathered}
\bar{\bm{x}}_j=(1-\alpha)\bar{\bm{x}}_{j-1}+\alpha\bm{x}_j \\
\hat{G}_j(\triangle t)=(1-\alpha)\left[\hat{G}_{j-1}+\alpha({\bm{x}_j}-\bar{\bm{x}}_{j})({\bm{x}_{j-1}}-\bar{\bm{x}}_{j-1})^T\right] \\
\hat{C}_j^{-1}=\frac{1}{(1-\alpha)}\left[\hat{C}_{j-1}^{-1}-\alpha\frac{\hat{C}_{j-1}^{-1}\bm{z}_j\bm{z}_j^T\hat{C}_{j-1}^{-1}}{1+\alpha(\bm{z}_j^T\hat{C}_{j-1}^{-1}\bm{z}_j-1)}\right]
\end{gathered}
\end{equation}
\normalsize
where $\bm{z}_j=(\bm{x}_j-\bar{\bm{x}}_{j-1})$. 
\item b) Compute the dynamic system state matrix:
\small
\begin{equation}
A_j=\frac{1}{\triangle t}\log\left[\hat{G}_{j}(\triangle t)\hat{C}_j^{-1}\right]\label{eq:algorithm-3}
\end{equation}
\normalsize
\end{itemize}
\subsection{Implementation Issues}

Like other measurement-based methods, the performance of the proposed recursive algorithm may be affected by parameter values such as the initialization window length and the smoothing factor $\alpha$.

Suppose the system is around the steady-state operating point, the larger initialization window will give more accurate estimations. Given the study in \cite{Wangxz:2017TRWS}, the estimation error of the algorithm in \cite{Wangxz:2017TRWS} does not decrease substantially as the window length increases beyond 200s for the IEEE 39-bus system. Hence, we use 200s initialization window for the same 39-bus system in this paper which shows reasonably good accuracy (See Fig. \ref{windowlength} for more details). In practice, the selection of initial data length may be decided by the off-line study of the system of interests and the PMU data quality.

Another critical factor is the smoothing factor $0\leq\alpha\leq1$. With $\alpha=\frac{1}{N}$, we assume the system operates in steady state, and hence all observations have the same weight independent on the time of occurrence; otherwise, a fluctuation is accepted such that the new observations may have a higher weight than the old ones which is indeed exponential smoothing, and the observation frame is approximately $\frac{1}{\alpha}$. Values of $\alpha$ close to 1 have a less smoothing effect and give greater weight to recent observations, while values of $\alpha$ closer to 0 have a greater smoothing effect and are less responsive to recent changes. In the limiting case with $\alpha=1$, the output is just the current observation. Generally, there is no reliable way of choosing $\alpha$, and it is mostly decided using heuristic methods \cite{Ross:1995}. In practice, $\alpha$ may be chosen adaptively based on the operating condition of the system. For instance, if the system is in steady state, let $\alpha=\frac{1}{N}$; if a contingency happens such that the system jumps to another operation state, choose a larger $\alpha$ to forget old samples more quickly and gradually decrease it until it reaches $\frac{1}{N}$. 

In this paper, the adaptive exponential smoothing shown in Algorithm \ref{adaptivesmoothing} is applied and integrated into the proposed algorithm in Section \ref{recursive method}. Particularly, $\beta$ and the change rate $w$ are two tunable parameters that affect how fast the algorithm could provide accurate estimation under a sudden change of the system, and describe the trade-off between the tracking capability under a sudden change and the accuracy of the estimation in steady state. 

\begin{algorithm}
\begin{algorithmic}
\caption{Adaptively choosing the smoothing factor $\alpha$.}
\label{adaptivesmoothing}
\State {\textbf{Given} $N,\Delta t,\beta,w,\hat{\bm{x}}_{j-1},\hat{C}_{j-1},\hat{G}_{j-1}(\Delta t),j_{c}$,flag,keep$\leftarrow$1.}
\While{keep}
    \State {\textbf{Read} current PMU measurements $\bm{x}_{j}$.}
    \If{The system experiences a sudden change at ${j}$.}
        \State {$j_{c}\leftarrow j$, flag $\leftarrow$ 1.}
    \EndIf
    \If{flag$=1$}
        \State{$\alpha \leftarrow \max\{\frac{1}{\beta+(j-j_{c})w}, \frac{1}{N}\}$. \hspace{0.1cm} \textit{\% In transient state}}
    \Else
        \State{$\alpha \leftarrow \frac{1}{N}$. \hspace{2.75cm} \textit{\% In steady state}}
    \EndIf
    \State{\textbf{Update} $\hat{\bm{x}}_{j},\hat{G}_{j}(\Delta t)$, and $\hat{C}_{j}^{-1}$ by equation \eqref{eq:recurformula}.}
    \State{$j \leftarrow j+1$.}
\EndWhile
\end{algorithmic}
\end{algorithm}
%


%
\section{Case Studies}\label{casestudy}
\subsection{Numerical Example I: Estimating the Dynamic System State Matrix}\label{sectionexampleI}

First, we use a small system---the standard Western System Coordinating Council (WSCC) 3-generator, 9-bus system model to validate the proposed method, i.e., the accuracy of (\ref{samplesystemmatrix}). 
The stochastic dynamic model integrating the load variations in the center-of-inertia (COI) formulation can be represented as \cite{Wangxz:2017TRWS}, \cite{Crow:2013}:
\small
\begin{equation}
\begin{aligned}
\label{9bus-1}
\dot{\tilde{\delta}}_i&=\tilde{\omega}_i \\
M_i\dot{\tilde{\omega}}_i&=P_{m_i}-P_{e_i}-\frac{M_i}{M_T}P_{coi}-D_i\tilde{\omega}_i-E_i^2G_{ii}\sigma_i\xi_i \\
&+\frac{M_i}{M_T}\sum_{j=1}^3{E_j^2 G_{jj} \sigma_j \xi_j}
\end{aligned}
\end{equation}
\normalsize
where
\small
\begin{equation}
\begin{gathered}
M_T=\sum_{j=1}^{3}M_j \\
\delta_0=\frac{1}{M_T}\sum_{j=1}^{3}M_j\delta_j,\enskip \omega_0=\frac{1}{M_T}\sum_{j=1}^{3}M_j\omega_j \\
\tilde{\delta}_i=\delta_i-\delta_0,\enskip \tilde{\omega}_i=\omega_i-\omega_0,\enskip i=1,2,3
\end{gathered}
\end{equation}
\normalsize
\small
\begin{equation}
\begin{gathered}
P_{e_i}=\sum_{j=1}^{3}E_iE_j(G_{ij}\cos(\tilde{\delta}_i-\tilde{\delta}_j)+B_{ij}\sin(\tilde{\delta}_i-\tilde{\delta}_j)) \\
P_{coi}=\sum_{i=1}^{3}(P_{m_i}-P_{e_i}),\enskip i=1,2,3
\end{gathered}
\end{equation}
\normalsize

The model-based system state matrix is as follows:
\small
\begin{equation}
A=\left[\begin{array}{cc|cc}
0&0&1&0 \\
0&0&0&1 \\
\hline
\multicolumn{2}{c}{\multirow{2}{*}{$-M^{-1}\left(\frac{\partial\bm{P_e}}{\partial\bm{\tilde{\delta}}}\right)_{coi}$}}\vline&-\frac{D_1}{M_1}&0 \\
&&0&-\frac{D_2}{M_2}\end{array}\right]
\label{A}
\end{equation}
\normalsize
where
\small
\begin{equation}
\left(\frac{\partial\bm{P_e}}{\partial\bm{\tilde{\delta}}}\right)_{coi}=\frac{\partial\bm{P_e}}{\partial\bm{\tilde{\delta}}}+M\frac{1}{M_T}\frac{\partial P_{coi}}{\partial\bm{\tilde{\delta}}}
\end{equation}
\normalsize
The diagonal element and off-diagonal element of $\left(\frac{\partial\bm{P_e}}{\partial\bm{\tilde{\delta}}}\right)_{coi}$ can be calculated by
\small
\begin{equation}
\label{dpeddcoi}
\begin{aligned}
\left[\left(\frac{\partial\bm{P_e}}{\partial\bm{\tilde{\delta}}}\right)_{coi}\right]_{ii}&=\sum_{k\not=i}E_iE_k[-G_{ik}\sin(\tilde{\delta}_i-\tilde{\delta}_k)+B_{ik}\cos(\tilde{\delta}_i-\tilde{\delta}_k)] \\
&+2\frac{M_i}{M_T}\sum_{k\not=i}E_iE_kG_{ik}\sin(\tilde{\delta}_i-\tilde{\delta}_k) \\
\left[\left(\frac{\partial\bm{P_e}}{\partial\bm{\tilde{\delta}}}\right)_{coi}\right]_{ij}&=E_iE_j[G_{ij}\sin(\tilde{\delta}_i-\tilde{\delta}_j)-B_{ij}\cos(\tilde{\delta}_i-\tilde{\delta}_j)] \\
&+2\frac{M_i}{M_T}\sum_{k\not=i}E_iE_kG_{ik}\sin(\tilde{\delta}_i-\tilde{\delta}_k)
\end{aligned}
\end{equation}
\normalsize

Note that the conventional model-based method of estimating $(\frac{\partial\bm{P_e}}{\partial\bm{\tilde{\delta}}})_{coi}$ and thus $A$ heavily depends on the network topology and parameter values embedded in the admittance matrix $Y$. In contrast, the proposed technique, i.e., (\ref{samplesystemmatrix}), is purely measurement-based, completely independent of the network model and the generator parameters.

\begin{figure}[ht]
\centering
\includegraphics[width=0.45\textwidth,keepaspectratio=true,angle=0]{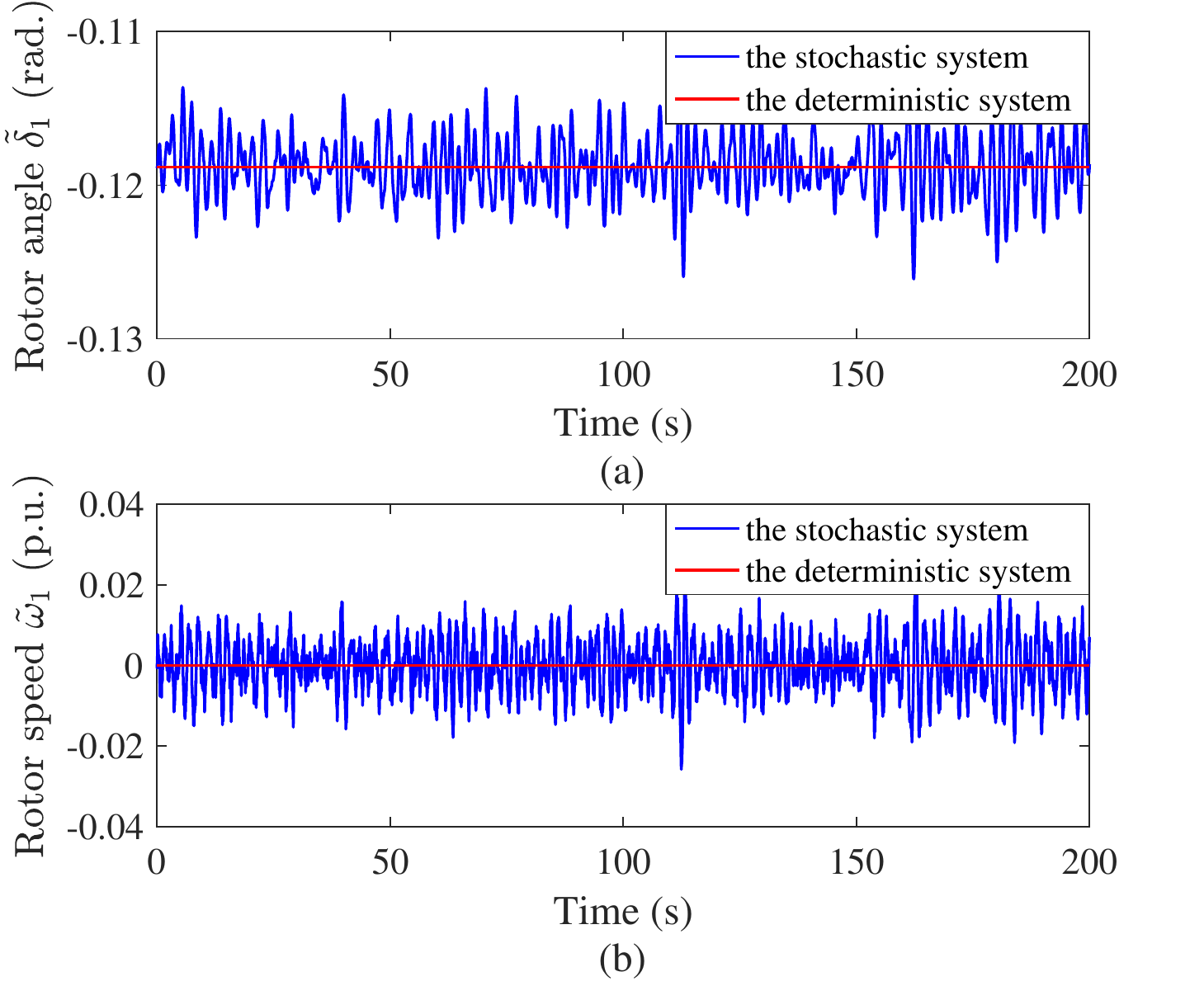}
\caption{Trajectories of $\tilde{\delta}_1$ and $\tilde{\omega}_1$ in the 9-bus system when the standard deviation of load variations are set to 0.01. (a) Trajectory of $\tilde{\delta}_1$ on $[0s,200s]$. (b) Trajectory of $\tilde{\omega}_1$ on $[0s,200s]$.}
\label{9bus}
\end{figure}

Assuming $\sigma_1=\sigma_2=0.01$ p.u. describing the standard deviations of the load variations, Fig. \ref{9bus} presents the emulated $\tilde{\delta}_1$ and $\tilde{\omega}_1$ estimated from the PMU data with a sampling rate of 50 Hz, from which we see that the system's trajectories are oscillating around their steady-state solutions (denoted by the red lines) because of the stochastic load variations. 

If there is no stochastic load variation, i.e., $\sigma_1=\sigma_2=0$ p.u., the dynamic system state matrix $A$ would be a constant matrix calculated from (\ref{A})-(\ref{dpeddcoi}) using the steady-state operating point:
\small
\begin{equation}
A=\left[\begin{array}{cccc}0&0&1&0\\0&0&0&1\\-12.84&-1.98&-1&0\\-8.25&-14.98&0&-1\end{array}\right]\label{9busA}
\end{equation}
\normalsize

To validate the proposed algorithm, we will show that the dynamic system state matrix estimated by the proposed algorithm is close to the deterministic system state matrix (\ref{9busA}). 
To this end, we first use the $200$s data to estimate the sample covariance matrix and the sample one-lag correlation matrix by (\ref{samplemean})-(\ref{samplecorrelatoin}):
\small
\begin{equation}
\hat{C}_{\bm{x}\bm{x}}=10^{-3}\times\left[\begin{array}{cccc} 0.0042 & -0.0059 &-0.0004 &-0.00003\\-0.0059&0.0101&0.0012&-0.0011\\-0.0004&0.0012&0.0431&-0.0569\\-0.00003&-0.0011&-0.0569&0.1065 \end{array}\right]\nonumber
\end{equation}
\normalsize
\small
\begin{equation}
\begin{aligned}
\hat{G}_{\bm{x}\bm{x}}(0.02)&=10^{-3}\times\\
&\left[\begin{array}{cccc} 0.0041&-0.0059&0.0004&-0.0012\\
-0.0059&0.0101&0.00003&0.0011\\
-0.0013&0.0023&0.0423&-0.0558\\
0.0011&-0.0032&-0.0560&0.1046 \end{array}\right] \nonumber
\label{Jacobiandistance}
\end{aligned}
\end{equation}
\normalsize
and then estimate the dynamic system state matrix $A$ by (\ref{samplesystemmatrix}):
\small
\begin{equation}
A^\star=\left[\begin{array}{cccc}0.1390&0.0211&1.0086&0.0003\\0.0702&0.1507&-0.0019&1.0082\\
-13.5099&-2.1961&-0.9437&-0.0576\\-7.7199&-15.0019&-0.0677&-0.9749\end{array}\right]\label{estimated9busA}
\end{equation}
\normalsize
\noindent where $^\star$ denotes the estimation by the proposed method. It is observed that the estimated dynamic system state matrix (\ref{estimated9busA}) is close to the deterministic matrix shown in (\ref{9busA}). Particularly, we use the normalized Frobenius norm to quantify the difference between the actual and estimated matrices:
\small
\begin{equation}
\frac{\|A-A^\star\|_F}{\|A\|_F}=4.25\%\label{Jacobiandistance} \end{equation}
\normalsize
where $\|\|_F$ denotes the Frobenius norm. Note that we only use 200s data for demonstration. However, the accuracy will improve as the sampling rate increases and/or the length of measurement data grows. There is always a trade-off between the computational effort and the accuracy.

In addition, we compare the accuracy of the proposed method with that of the method developed in \cite{Wangxz:2017TRWS}. The method in \cite{Wangxz:2017TRWS} are developed based on another important property of the stationary covariance matrix $C$ that $C$ satisfies the Lyapunov equation in ambient condition:
\begin{equation}
AC_{\bm{xx}}+C_{\bm{xx}}A^T=-BB^T
\end{equation}
After algebraic reduction, we have:
\begin{equation}
\frac{\partial{\bm{P_e}}}{\partial{\bm{\delta}}}=MC_{\bm{\omega\omega}}C^{-1}_{\bm{\delta}\bm{\delta}}+DC_{\bm{\delta}\bm{\omega}}C^{-1}_{\bm{\delta}\bm{\delta}}\label{eq:lyapunov}
\end{equation}
Assuming that the contributions from the matrix $C_{\bm{\delta\omega}}$ are negligible, it yields
$\frac{\partial{\bm{P_e}}}{\partial{\bm{\delta}}}\approx 
MC_{\bm{\omega\omega}}C^{-1}_{\bm{\delta}\bm{\delta}}$.
It is obvious that this method is not purely measurement-based since the dynamic system state matrix $\frac{\partial{\bm{P_e}}}{\partial{\bm{\delta}}}$ depends on the inertia constants $M$. If $A$ needs to be estimated, $D$ has to be known. 
In this numerical example, the estimation error defined in (\ref{Jacobiandistance}) is $3.73\%$ using (\ref{eq:lyapunov}), which is less than that of the proposed method. 
Nevertheless, since the method in this paper and the one in \cite{Wangxz:2017TRWS} are developed based on different properties of the stochastic dynamical system (\ref{eq:OU}), there is no direct and clear conclusion that which method should possess a higher accuracy, even though the method proposed in this paper requires less model information and is completely model-free.

\subsection{Numerical Example II: Detecting Topology Change and Identifying the Location}\label{subsectionexampleII}

In this example, we apply the \textbf{recursive} estimation method to a larger system---the IEEE 39-bus 10-generator test system to demonstrate that the proposed methodology is able to detect the topology change and more importantly, identify the corresponding location. This information is of great significance for power system operators to adopt corrective control strategies to fix the error and maintain the security of power grids.

\begin{figure}[!ht]
\centering
\includegraphics[width=0.49\textwidth,keepaspectratio=true,angle=0]{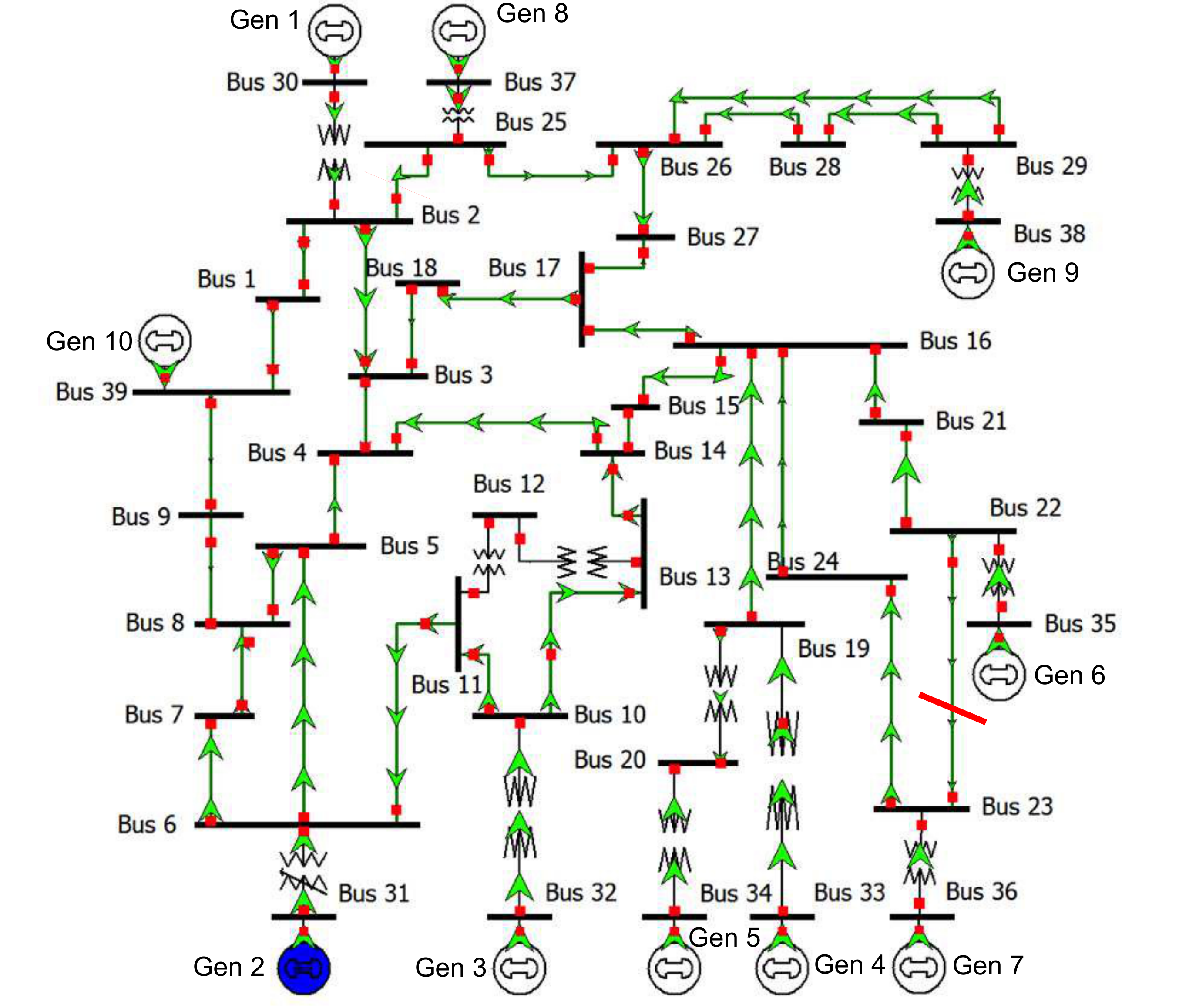}
\caption{The network topology of the IEEE 39-bus 10-generator system \cite{IEEE39bus}.} \label{ieee-39}
\end{figure}

\begin{figure}[!ht]
\centering
\includegraphics[width=0.45\textwidth,keepaspectratio=true,angle=0]{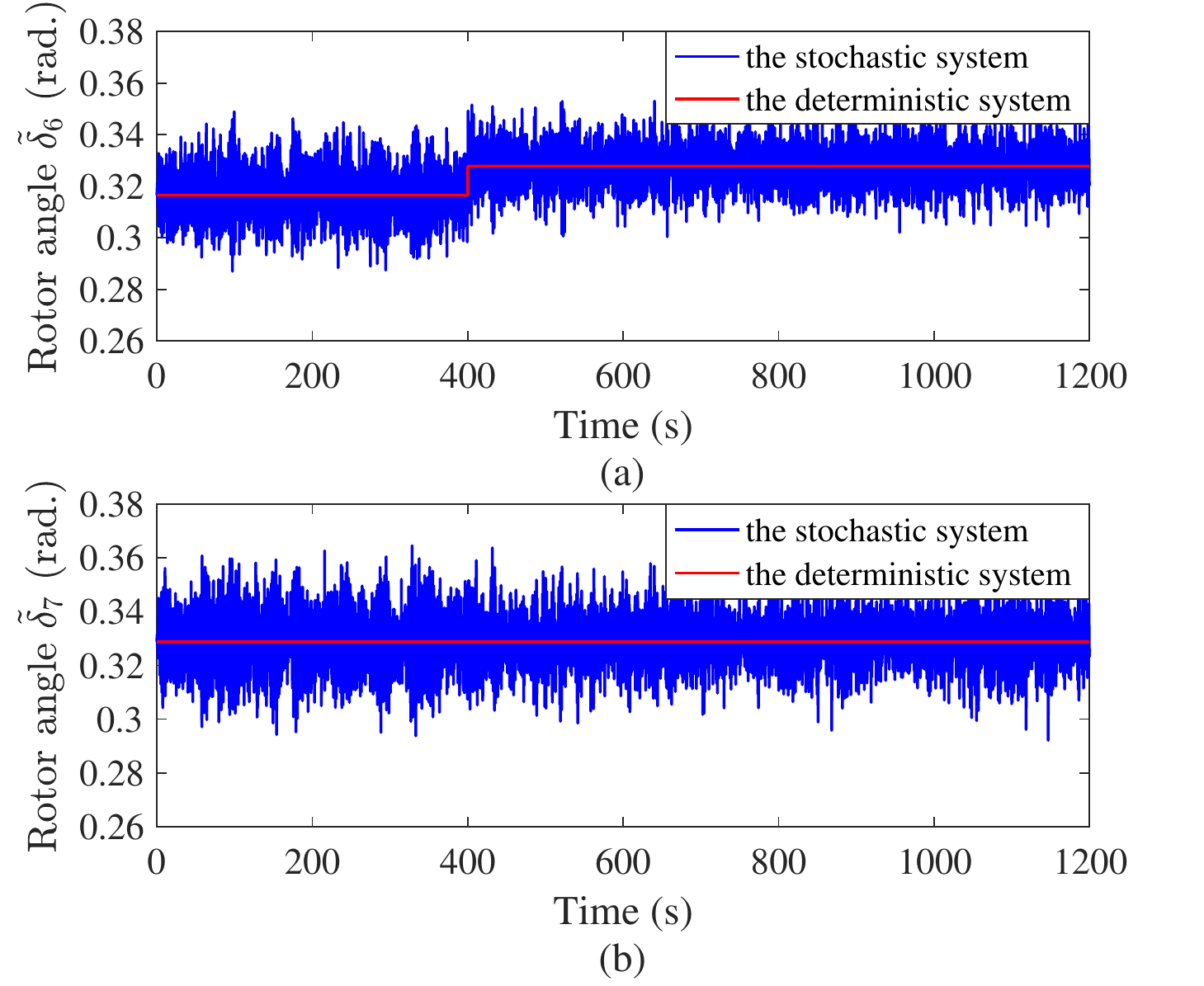}
\caption{Trajectories of rotor angles $\tilde{\delta}_6$ and $\tilde{\delta}_7$ in the 39-bus system when an undetected line tripping occurs at 400s. (a) Trajectory of $\tilde{\delta}_6$ on $[0s,1200s]$. (b) Trajectory of $\tilde{\delta}_7$ on $[0s,1200s]$.}
\label{39bus_topo}
\end{figure}

For this purpose, we assume that an undetectable line tripping between Bus 22 and Bus 23 occurs at 400s due to, for instance, communication errors. As shown in Fig. \ref{ieee-39}, the topology change location is close to Generator 6 and 7. The trajectories of the rotor angles of Generator 6 and 7 are shown in Fig. \ref{39bus_topo}, from which we see that there is an abrupt change in $\tilde{\delta}_6$ at 400s. Nevertheless, the post-fault system still maintains the stability. 

Suppose the change of $\bm{\delta}$ and $\bm{\omega}$ can be detected by the PMU measurements, yet the network topology change is not detected. Denote the dynamic system state matrix estimated by the model-based method without accounting for the topology change as $A^\diamond$, it is found that the normalized distance between $A^\diamond$ and the true dynamic system state matrix $A$ after the change is $17.90\%$. The big discrepancy is due to the incorrectly assumed network model. In contrast, the normalized distance between the dynamic system state matrix $A^\star$ estimated by the proposed method and the true one is equal to $2.72\%$.
Particularly, the data on [410s 1200s] is used to estimate the covariance matrix and the one-lag correlation matrix. It is obvious that the proposed method provides more accurate estimation than the model-based method under the occurrence of undetectable topology change. 
In addition, the big difference between the model-based and the measurement-based matrices indicates that the assumed system model is inaccurate and identifying the location of the error is stringent. It will be shown next that the proposed recursive matrix estimation algorithm can also help identify the location of the topological error.

\begin{figure}[!ht]
\centering
\includegraphics[width=0.45\textwidth,keepaspectratio=true,angle=0]{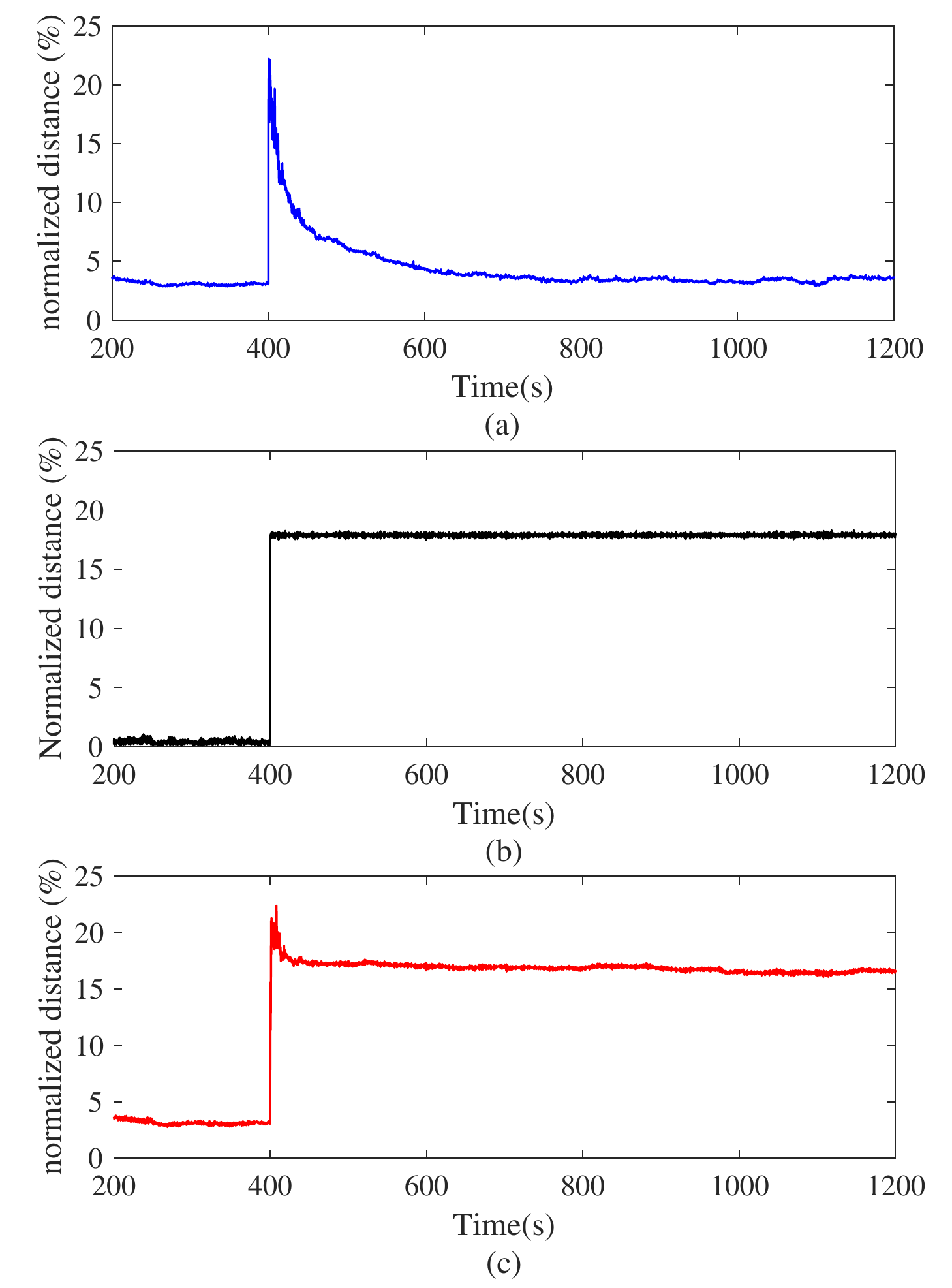}
\caption{The normalized distance between matrices with respect to time under the undetected topology change. (a) The distance between the true matrix and the one estimated by the proposed method. (b) The distance between the true matrix and the one estimated by the model-based method. (c) The distance between the matrix estimated by the proposed method and the one by the model-based method.}
\label{recursiveerror}
\end{figure}

\begin{figure}[!ht]
\centering
\includegraphics[width=0.45\textwidth,keepaspectratio=true,angle=0]{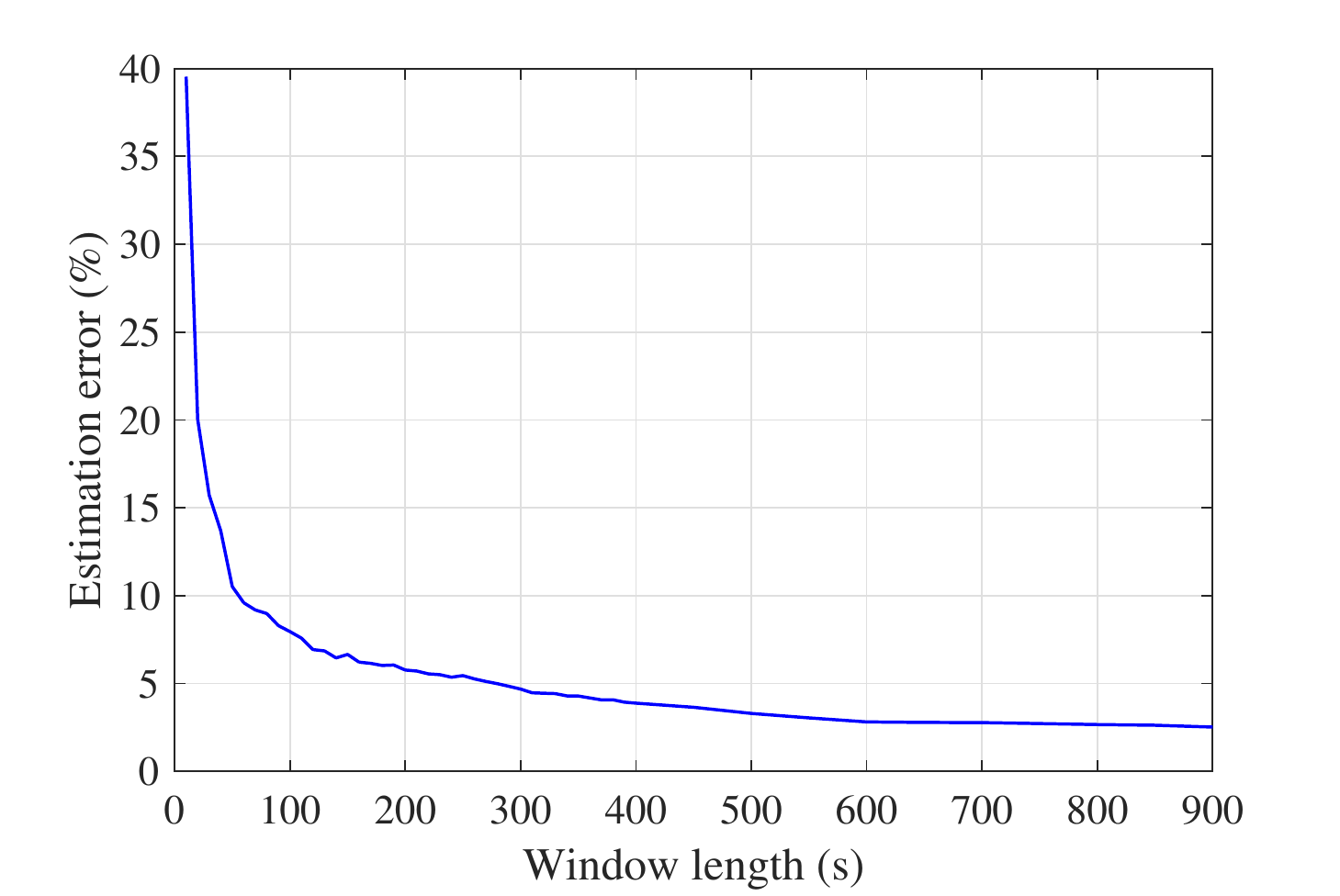}
\caption{The estimation error (\%) as the window length (seconds) changes from 10s to 900s for the 39-bus system.}
\label{windowlength}
\end{figure}

To implement the recursive algorithm, we set the initial window length to be 200s, i.e., $N=10000$ for a sampling rate of $50$ Hz; $\beta=200$ and $w=2$ in the adaptive algorithm for $\alpha$ (Algorithm \ref{adaptivesmoothing}). A small $\beta$ means that a large smooth factor $\alpha$ is applied to quickly forget the previous samples and $w$ describes a gradual decrease in $\alpha$ until it reaches its steady-state value $\frac{1}{N}$ or when the system returns to the steady state, after which $\alpha$ is reset to be $\frac{1}{N}$.
The normalized Frobenius norm between the estimated dynamic system state matrix and the true one is shown in Fig. \ref{recursiveerror}(a), from which we see that the proposed method is able to provide an accurate estimation for the dynamic system state matrix both before and after the line tripping. 
It should be noted that the normalized distance shown in Fig. \ref{recursiveerror} (a) by the recursive algorithm is higher than $2.72\%$ when the data on [410s 1200s] is used. Such difference lies in the fact that different window lengths are applied, i.e., 790s vs. 200s. The impact of window length will be further discussed in Fig. \ref{windowlength}.
Besides, the errors in the transient period 400s-600s are due to the fact that it takes some time for the recursively estimated covariance matrix, the one-lag correlation matrix, and thus the system state matrix to settle down to their new values corresponding to the new steady state after the topology change. In addition, the normalized Frobenius norm between the estimated dynamic system state matrix by the model-based method (i.e., without detecting the topology change) and the true one is shown in Fig. \ref{recursiveerror}(b), showing that the model-based method matrix becomes inaccurate after the undetected topology change. 
Given the fact that the true dynamic system state matrix is typically unknown in practice, we also compare the normalized distance between the dynamic system state matrix estimated by the proposed recursive method and the one obtained from the conventional model-based method. 
As shown in Fig. \ref{recursiveerror}(c), the two estimated matrices are close to each other with a difference around 3\% (both the model-based method and the proposed measurement-based method work well), yet they deviate from each other after the line tripping due to the incorrectly assumed network model used by the model-based method. 
By monitoring the difference between the two estimated matrices, power system operators can quickly detect that an error occurs in the assumed network model at 400s and locating the error is imperative. 

To justify that a window length of 200s is reasonable, we compare the accuracy of the estimation when the window length changes from 10s to 900s as shown in Fig. \ref{windowlength}. It can be seen that the estimation accuracy does not improve dramatically when the window length goes beyond 200s. Considering the compromise between accuracy and efficiency, we choose $200$s to be the length of the initial window and the moving window in the recursive algorithm. 

Regarding the efficiency, the proposed recursive algorithm is not computationally expensive since it involves only basic arithmetic operations as seen from (\ref{eq:algorithm-1})-(\ref{eq:algorithm-3}). In our simulation study, the average time consumed for each time instant $j$ in \textbf{Step 2} where the main computational effort lies in is $1.8592\times10^{-4}$s for the 3-generator system (\textit{Numerical Example I in Section \ref{sectionexampleI}}), and $8.7911\times10^{-4}$s for this 10-generator system. Note that there is always one generator being the reference. It can be observed that the time consumption grows approximately linearly with the number of generators. The efficiency can be further improved to conduct the recursive estimation less often in the steady state, whereas more often after the system's evolution due to, for instance, topology change. 

\begin{figure}[!ht]
\centering
\includegraphics[width=0.45\textwidth,keepaspectratio=true,angle=0]{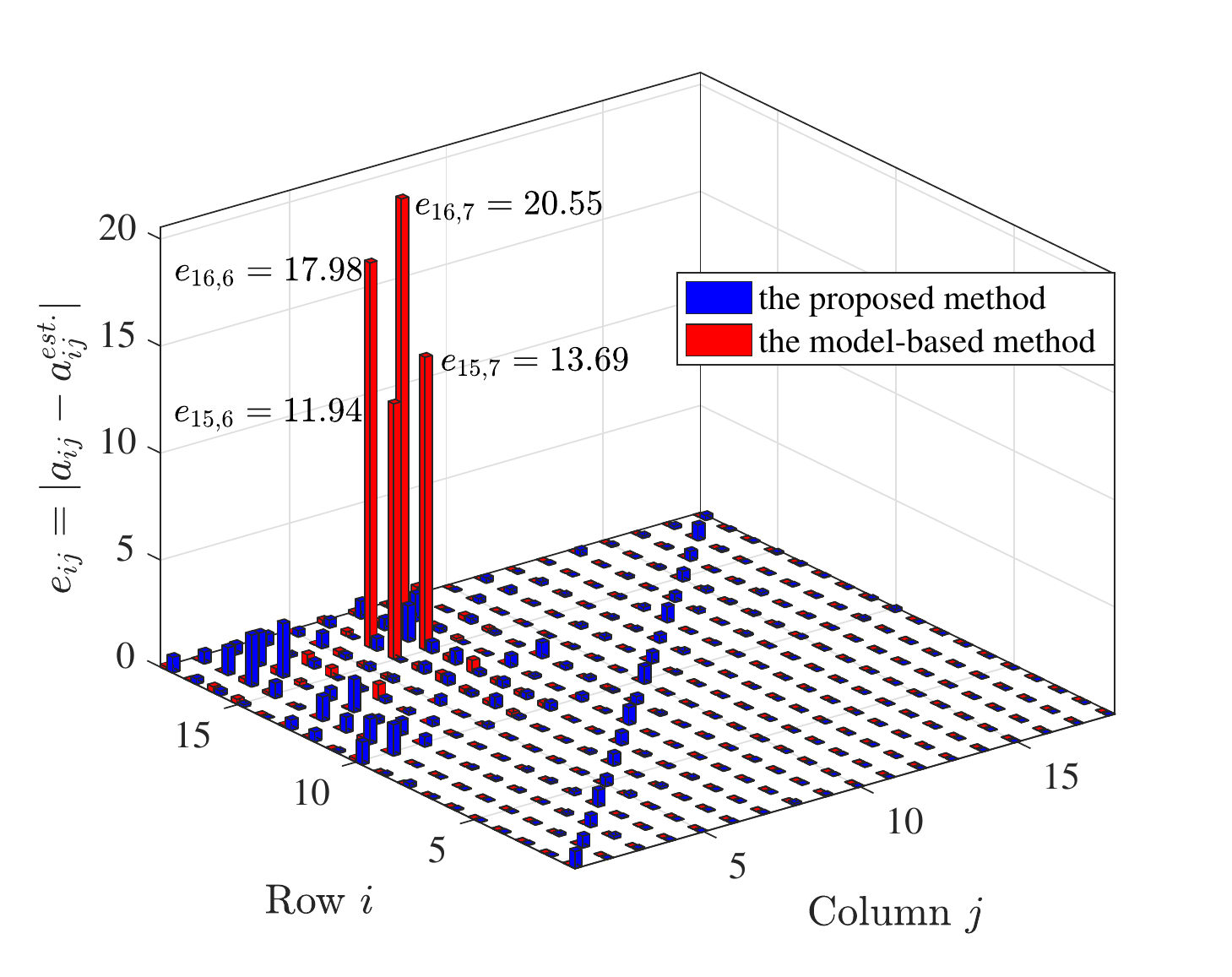}
\caption{The distance between each component of the true dynamic system state matrix and the estimated matrices. It is evident that the proposed method provides a more accurate estimation than the model-based method.}
\label{matrixsurf}
\end{figure}

\begin{figure}[!ht]
\centering
\includegraphics[width=0.45\textwidth,keepaspectratio=true,angle=0]{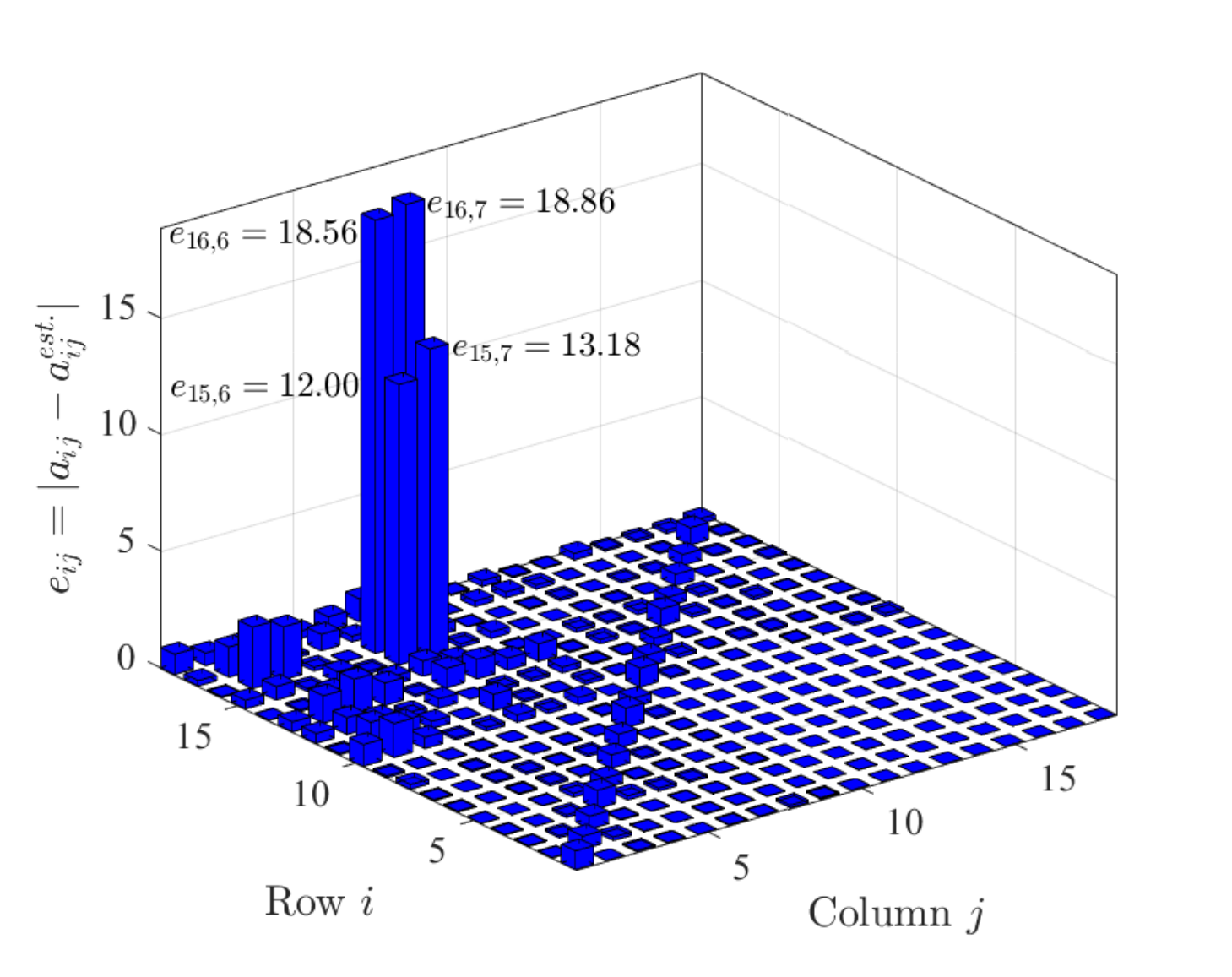}
\caption{The distance between each component of the dynamic system state matrices estimated by the model-based method and by the proposed measurement-based method, respectively. Large discrepancies indicate that there may be undetected topology changes.}
\label{matrixsurfdiff}
\end{figure}

To better visualize the difference between the true matrix, the estimated matrix by the proposed method, and the estimated matrix by the model-based method, we use the 3D bar graph to show the difference between each component of the matrices. Denote $A^\diamond(t)$ as the matrix estimated by the model-based method using the measurement at $t$, and $A^\star(t)$ as the matrix estimated by the recursive measurement-based method at time $t$, then Fig. \ref{matrixsurf} presents the difference between $|a_{ij}-a_{ij}^\diamond(1200)|$ and $|a_{ij}-a_{ij}^\star(1200)|$, where $a_{ij}\in{A}$ --- the true system matrix, $a_{ij}(1200)^\diamond\in{A^\diamond(1200)}$, and $a_{ij}(1200)^\star\in{A^\star(1200)}$. Furthermore, Fig. \ref{matrixsurfdiff} directly shows the difference between the component of the two estimated matrices which we can acquire online, i.e., $|a_{ij}^\diamond(1200)$-$a_{ij}^\star(1200)|$. It is evident from the figures that the proposed measurement-based method provides more accurate estimation than the model-based method. More importantly, the proposed method provides important insights regarding the location of the error. Since the largest discrepancies between the two estimated matrices (Fig. \ref{matrixsurfdiff}) occur in elements $a_{15,6}$, $a_{15,7}$, $a_{16,6}$, and $a_{16,7}$ corresponding to the ${\frac{\partial P_e}{\partial \delta}}$ of Generator 6 and 7, it indicates that the topology change is electrically close to Generator 6 and 7, which aligns exactly with the fact.

The results of the above two cases entail that the proposed recursive measurement-based method is able to provide a good estimation for the dynamic system state matrix in near real-time by utilizing the inherent statistical properties of the stochastic system. Moreover, the method may help detect the topological errors in the assumed network model, and more importantly, identify their sources.

\subsection{Numerical Example III: Identifying the Location of a Topology Change with Missing PMUs}

It is assumed that all generators are monitored by PMUs when developing the methodology, which seems to be too strict. In this section, we relieve this assumption and consider a more realistic situation where some PMUs are missing.

Assuming the PMUs are only available for Generators $K$, where $K=\{m_1,...,m_p\}$, then we can still accurately estimate the sample mean $\bar{x}_i$ where $i\in K$, and the sub-matrices $\hat{C}_{ij}$ and $\hat{G}_{ij}$, where $i, j\in K$, of the sample covariance matrix $\hat{C}$ and the sample correlation matrix $\hat{G}$, respectively. As a result, by (\ref{systemmatrix}), we can accurately achieve a sub-matrix $A_{ij}$, where $i, j\in K$, of the dynamic system state matrix $A$. More importantly, the estimated sub-matrix can still facilitate the detection and location of the topology change.

\begin{figure}[!ht]
\centering
\includegraphics[width=0.45\textwidth,keepaspectratio=true,angle=0]{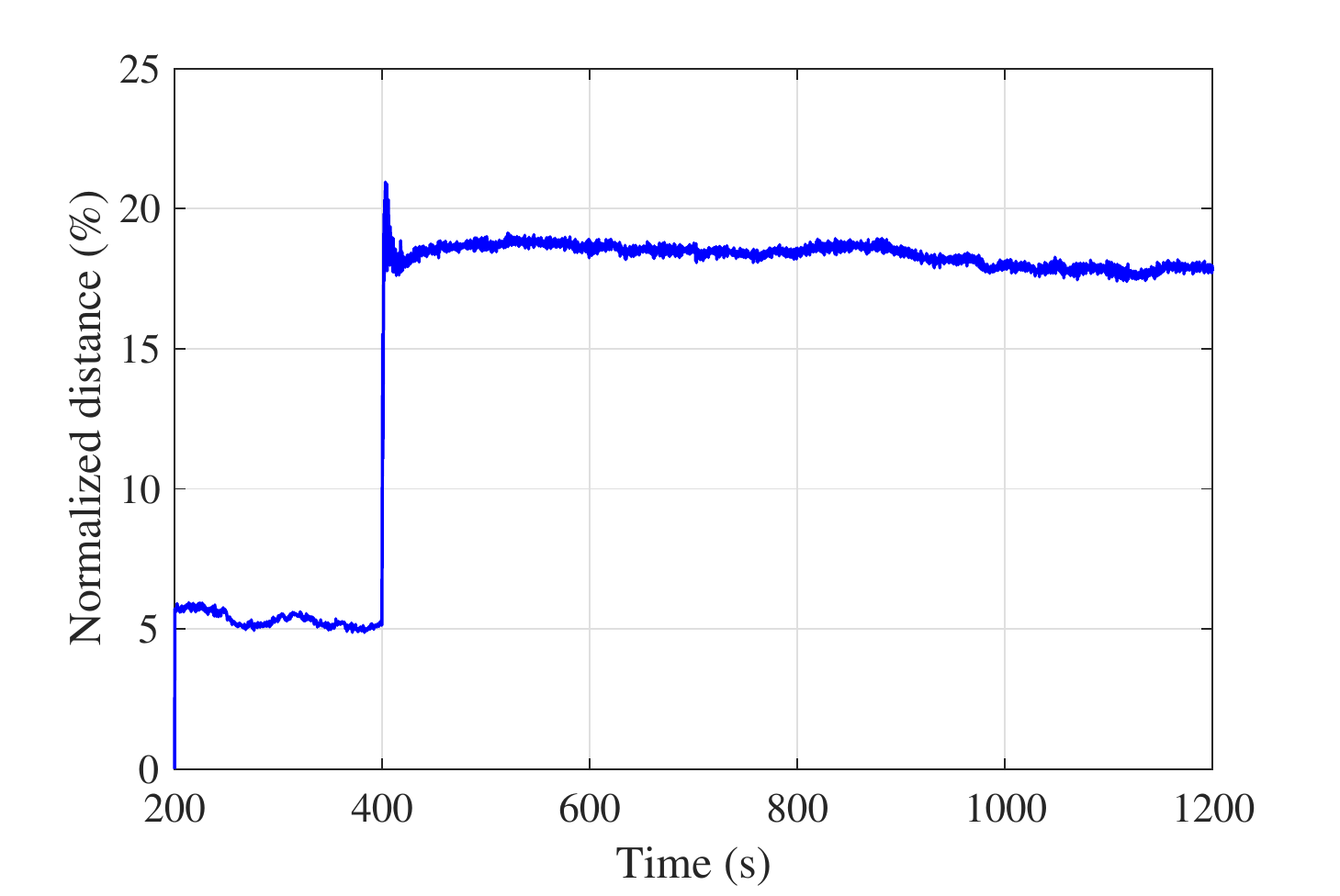}
\caption{The distance between the dynamic system state matrix estimated by the proposed method and the one by the model-based method.}
\label{recursiveerror_ests_lost}
\end{figure}

\begin{figure}[!ht]
\centering
\includegraphics[width=0.45\textwidth,keepaspectratio=true,angle=0]{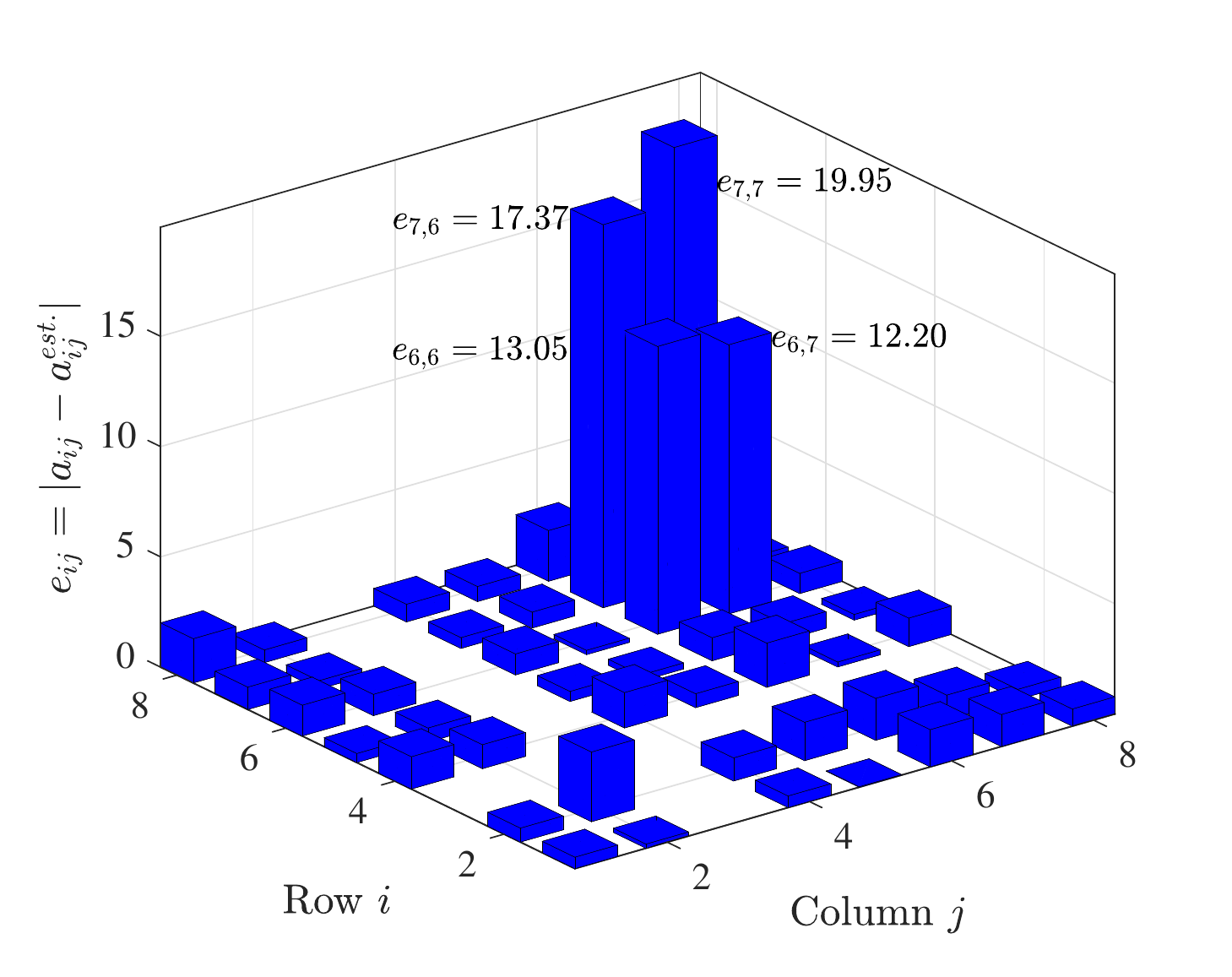}
\caption{The distance between each component of the sub-matrix estimated by the proposed measurement-based method and the one by the model-based method. The largest discrepancies occur in generator 6 and 7, indicating that the error is electrically close to these two generators.}
\label{matrixsurfdiff_lost}
\end{figure}

To show this, we reconsider the IEEE 39-bus 10-generator system in Section \ref{subsectionexampleII}. Assuming the PMUs at Gen 3 and Gen 9 are missing, i.e., $K=\{1,2,4,5,6,7,8\}$ and $\delta_{10}$ is the reference, we keep estimating the difference between the two sub-matrices recursively by the proposed measurement-based method and by the model-based method, respectively. The Frobenius norm of the difference with respect to the time is presented in Fig. \ref{recursiveerror_ests_lost}, from which it can be concluded that the error in the network model occurring at 400s was not well detected. To locate the error, we further draw the 3D bar graph to spot the difference between each component of the two sub-matrices at a certain time step after the fault, say 1200s, as shown in Fig. \ref{matrixsurfdiff_lost}. Note that only the submatrix ${-M^{-1}(\frac{\partial\bm{P_e}}{\partial\bm{\tilde{\delta}}})_{coi}}$ in \eqref{A} is presented since it possesses the largest discrepancies as revealed in Fig. \ref{matrixsurf}-\ref{matrixsurfdiff}. Although the estimation for $A_{3j}$, $A_{i3}$, $A_{9j}$ and $A_{i9}$ are missing, we still can observe that the largest discrepancies occur in Gen 6 and 7, indicating that the error is electrically close to these two generators which well aligns with the fact.

It is worth mentioning that pinpointing the exact location of the topology error for a large system may require additional information since the proposed method relies on the reduced Y-matrix for the detection, in which only the internal generator buses are retained. Nevertheless, the proposed method may help narrow the range of topology change and identify the generators that are electrically close to the topology change without any knowledge of the network model.

\subsection{Impact of Measurement Noise}

\begin{figure}[!ht]
\centering
\includegraphics[width=0.45\textwidth,keepaspectratio=true,angle=0]{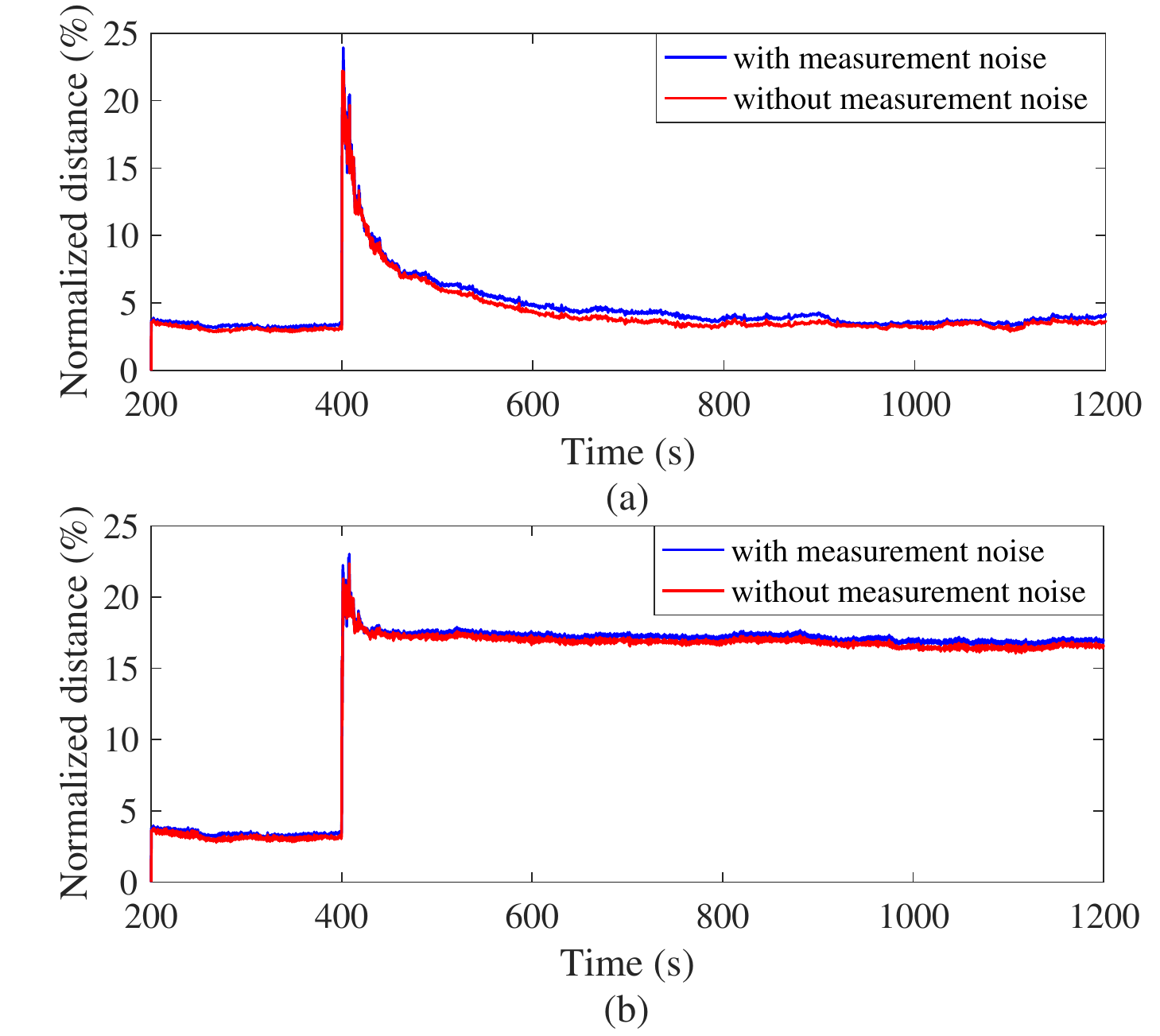}
\caption{The normalized distance between matrices under the undetected topology change when measurement noise is considered. (a) The distance between the estimated matrices by the proposed method and the true matrices. (b) The distance between the dynamic system state matrices estimated by the proposed method and the ones by the model-based method.}
\label{recursiveerror_mn}
\end{figure}

Similar to all the measurement-based methods, the noise may affect the accuracy of the proposed method. To investigate this issue, we added a measurement noise with a standard deviation $10^{-3}$ to both $\bm{\delta}$ and $\bm{\omega}$ of Numerical Example II according to the IEEE Standards \cite{IEEEStandard_amd}. Likewise, Fig. \ref{recursiveerror_mn}(a) shows the recursive estimation of the normalized distance between the estimated dynamic system state matrix by the proposed method and the true one; Fig. \ref{recursiveerror_mn}(b) presents the normalized distance between the matrix estimated by the proposed measurement-based method and the one by the model-based method. For comparison purpose, the results with no measurement noise are also plotted. It can be found that the impact of measurement noise on the accuracy of the method is minor since only a slight increase in the error is observed, which does not affect the detection of the topology error. 3D bar graph similar to Fig. \ref{matrixsurf}-\ref{matrixsurfdiff} can also be obtained, which provide essential information to locate the undetected topology change.
\subsection{Validation on a High-Order Model}

Despite the fact that the methodology is developed using the classical second-order generator model, it is necessary to test the performance of the methodology in a real-life setting in which higher-order models controlled by exciters are typical. To this end, we revisit the WSCC 3-generator, 9-bus system in Numerical Example I. Specifically, all generators are modeled as third-order models equipped by automatic voltage regulators. Assuming an undetected line tripping between Bus 7 and Bus 8 occurs at 300s, we apply the proposed recursive algorithm to estimate the dynamic system state matrix, detect the topology change and locate the error. Simulation has been carried out in PSAT-2.1.10 \cite{Milano:PSAT}.

\begin{figure}[!ht]
\centering
\includegraphics[width=0.45\textwidth,keepaspectratio=true,angle=0]{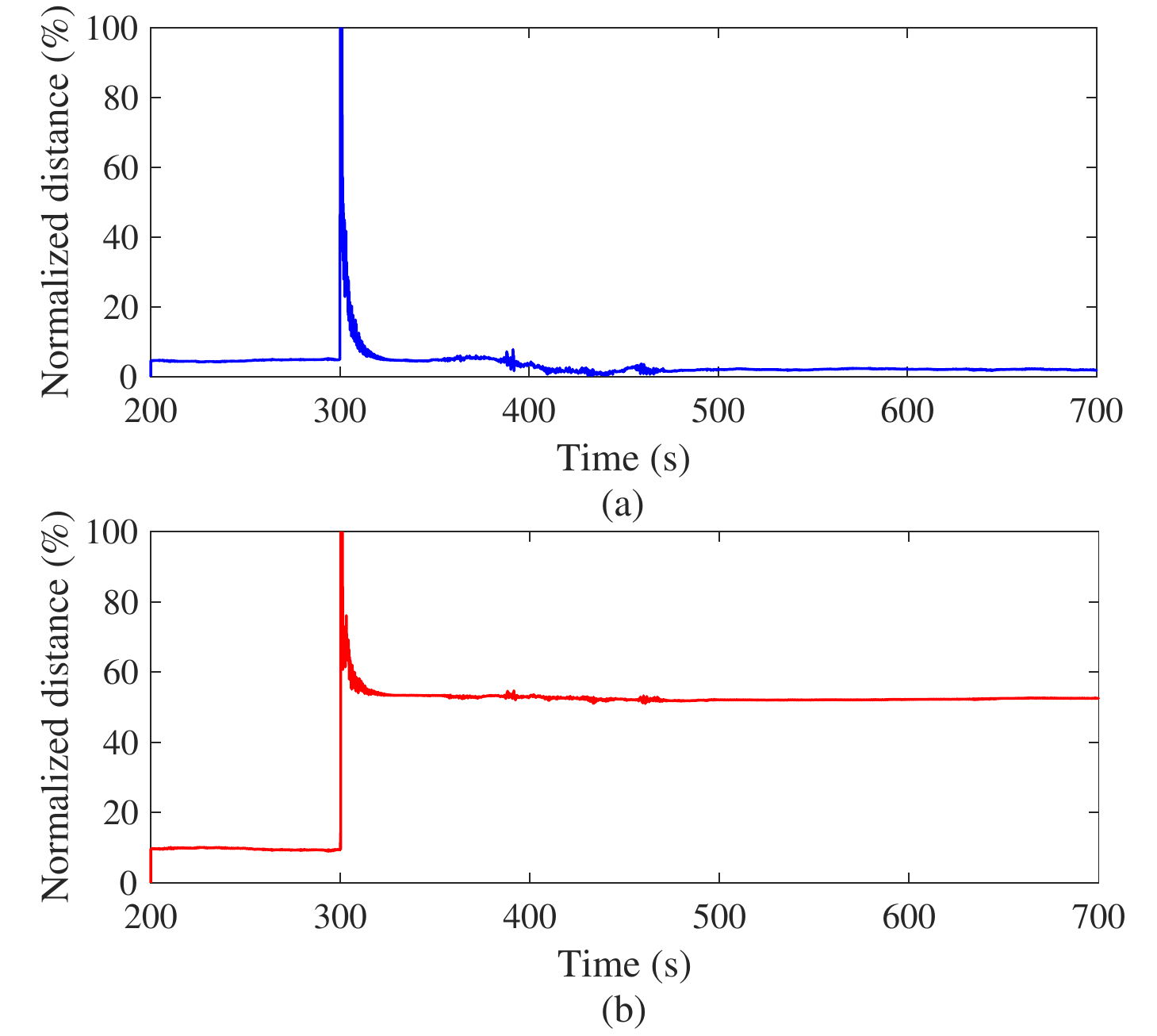}
\caption{The normalized distance between matrices for the 9-bus system under the undetected topology change, using third-order generator models equipped with exciters. (a) The distance between the true matrix and the one estimated by the proposed method. (b) The distance between the matrix estimated by the proposed method and the one by the model-based method.}
\label{recursiveerror_highorder}
\end{figure}

Particularly, the initial window length is set to be 200s; $\beta=10$ and $w=0.2$ are chosen for calculating the adaptive smooth factor $\alpha$. 
As mentioned earlier and revealed by Fig. \ref{matrixsurf}-\ref{matrixsurfdiff}, the submatrix ${-M^{-1}(\frac{\partial\bm{P_e}}{\partial\bm{\tilde{\delta}}})_{coi}}$ typically possesses the largest difference between matrices, which is therefore a good index to demonstrate the estimation accuracy, detect and locate topological change. The normalized difference between the true submatrix and the one estimated by the proposed method is shown in Fig. \ref{recursiveerror_highorder}(a), from which it can be seen that the proposed method can still provide accurate approximation regardless the implementation of third-order generator models with exciters. In addition, detecting topology change and identifying its location can still be realized by monitoring the difference between the submatrix estimated by the proposed measurement-based method and the one obtained from the model-based method, as illustrated in Fig. \ref{recursiveerror_highorder}(b).

\begin{figure}[!ht]
\centering
\includegraphics[width=0.45\textwidth,keepaspectratio=true,angle=0]{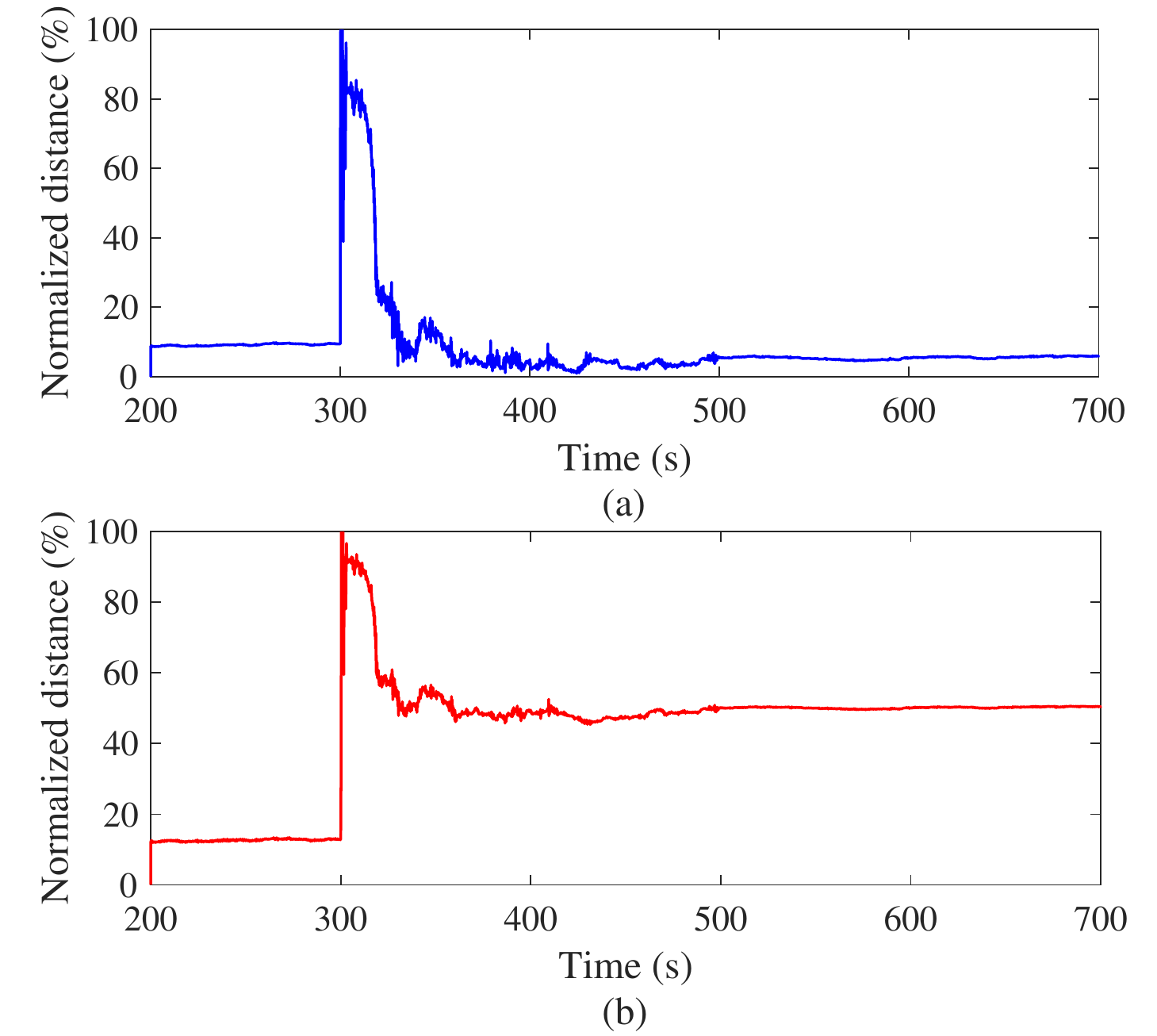}
\caption{The normalized distance between matrices for the 9-bus system under the undetected topology change, using fourth-order generator models with exciters and turbine governors. (a) The distance between the true matrix and the one estimated by the proposed method. (b) The distance between the matrix estimated by the proposed method and the one by the model-based method.}
\label{39_4order}
\end{figure}

Similar results can be obtained using the fourth-order model equipped by both automatic voltage regulators and turbine governors as shown in Fig. \ref{39_4order}. Although a small degradation in accuracy and convergence speed is noticed, the proposed method can still identify the topology change accurately and quickly, indicating the good feasibility of the proposed method in practice.
\subsection{Example IV: IEEE 145-bus System}
In order to validate the proposed method on large test systems, we applied it to the IEEE 145-bus dynamic stability test case. This system contains 50 generators and 453 branches. The power flow and dynamic parameters of this system are available online: \url{http://labs.ece.uw.edu/pstca/dyn50/pg_tcadd50.htm}.

The undetected topology change occurs between bus 117 and bus 118. The recursive algorithm is implemented with a window length of 200s. $\beta=200$, and $\omega=0.3$. Since the errors in the transient is too large, the normalized distance in the steady-state after the topology change have been zoomed in, showing that the proposed method provides more accurate estimation ($\approx8\%$) for the dynamic system state matrix than the model-based method ($\approx25\%$) under the undetectable topology change. 

\begin{figure}[!ht]
\centering
\includegraphics[width=0.45\textwidth,keepaspectratio=true,angle=0]{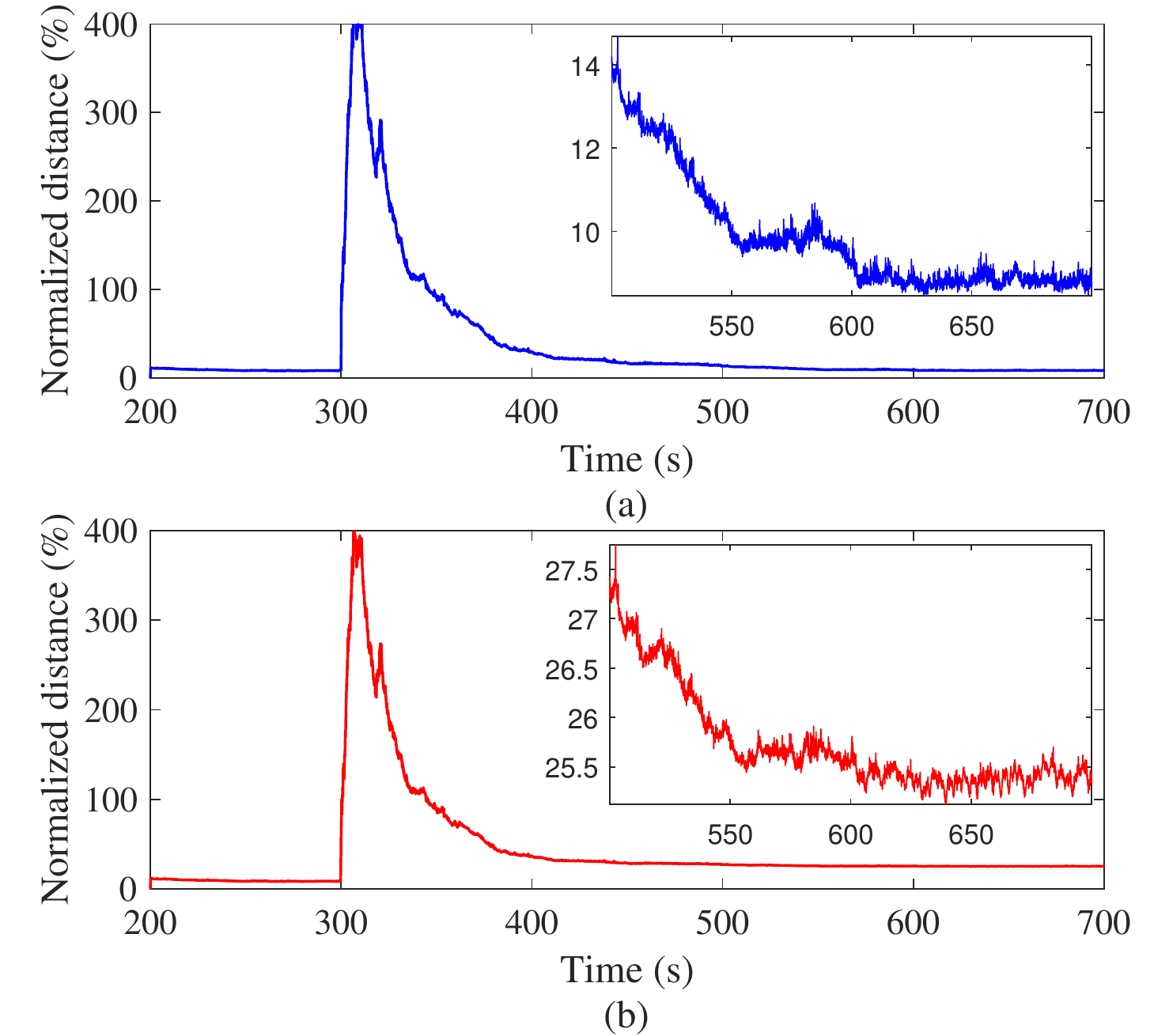}
\caption{The normalized distance between matrices for the 145-bus system under the undetected topology change. (a) The distance between the true matrix and the one estimated by the proposed method. (b) The distance between the matrix estimated by the proposed method and the one by the model-based method.}
\label{145_4order}
\end{figure}
%


%
\section{Conclusions and Perspectives}\label{sectionconclusion}

In this paper, we have proposed a novel recursive measurement-based method for estimating the dynamic system state matrix in near real-time. Particularly, the proposed method leverages the statistical properties of the stochastic dynamical system to extract significant model information out of PMU data. A recursive estimation algorithm has also been developed to enable online implementation. The simulation study has shown that the proposed recursive method can provide an accurate approximation for the dynamic system state matrix, and more importantly, detect severe discrepancies in the assumed network model and further identify their locations. Various implementation conditions including measurement noise, missing PMUs, and detailed generator models have been tested to demonstrate the feasibility and the accuracy of the proposed method in real-life applications.

In the future, we plan to explore diverse applications of the estimated dynamic system state matrix in online oscillation analysis, model validation, emergency power re-dispatch, and load shedding, etc. Besides, we intend to use actual, rather than emulated, PMU measurement to validate the proposed methodology in practice. 



\begin{thebibliography}{1}

\bibitem{Pai:1990}
P. W. Sauer and M. A. Pai, "Power system steady-state stability and the load-flow Jacobian," \textit{IEEE Trans. on Power Syst.}, vol. 5, no. 4, pp. 1374--1383, 1990.

\bibitem{PKundur94a}
P. Kundur, Power System Stability and Control, New York, NY, USA: McGraw Hill, 1994.

\bibitem{JHChow13a}
J. H. Chow, Power System Coherency and Model Reduction, New York, NY, USA: Springer-Verlag New York, 2013.

\bibitem{JDLRee10a}
J. D. L. Ree, V. Centeno, J. S. Thorp, and A. G. Phadke, "Synchronized phasor measurement applications in power systems," \textit{IEEE Trans. Smart Grid}, vol. 1, no. 1, pp. 20--27, 2010.

\bibitem{JZhao16a}
J. Zhao, G. Zhang, K. Das, G. N. Korres, N. M. Manousakis, A. K. Sinha, and Z. He, "Power system real-time monitoring by using {PMU}-based robust state estimation method," \textit{IEEE Trans. Smart Grid}, vol. 7, no. 1, pp. 300--309, 2016.

\bibitem{NZhou15a}
N. Zhou, D. Meng, Z. Huang, and G. Welch, "Dynamic state estimation of a synchronous machine using {PMU} data: a comparative study," \textit{IEEE Trans. Smart Grid}, vol. 6, no. 1, pp. 450--460, 2015.

\bibitem{CWLiu12a}
C. W. Liu, T. C. Lin, C. S. Yu, and J. Z. Yang, "A fault location technique for two-terminal multisection compound transmission lines using synchronized phasor measurements," \textit{IEEE Trans. Smart Grid}, vol. 3, no. 1, pp. 113--121, 2012.

\bibitem{MMajidi17a}
M. Majidi, M. E. Amoli, and M. S. Fadali, "A sparse-data-driven approach for fault location in transmission networks," \textit{IEEE Trans. Smart Grid}, vol. 8, no. 2, pp. 548--556, 2017.

\bibitem{JMa10a} 
J. Ma, P. Zhang, H. Fu, and Z. Dong, "Application of phasor measurement unit on locating disturbance source for low-frequency oscillation," \textit{IEEE Trans. Smart Grid}, vol. 1, no. 3, pp. 340--346, 2010.

\bibitem{SNabavi15a} 
S. Nabavi, J. Zhang, and A. Chakrabortty, "Distributed optimization algorithms for wide-area oscillation monitoring in power systems using interregional {PMU-PDC} architectures," \textit{IEEE Trans. Smart Grid}, vol. 6, no. 5, pp. 2529--2538, 2015.

\bibitem{Vittal:2012}
F. Ma and V. Vittal, "A hybrid dynamic equivalent using ANN-based boundary matching technique," \textit{IEEE Trans. on Power Syst.}, vol. 27, no. 3, pp. 1494--1502, Aug. 2012.

\bibitem{AChakrabortty11a}
A. Chakrabortty, J. H. Chow, and A. Salazar, "A measurement-based framework for dynamic equivalencing of large power systems using wide-area phasor measurements," \textit{IEEE Trans. Smart Grid}, vol. 2, no. 1, pp. 68--81, 2011.

\bibitem{AAHajnoroozi15a}
A. A. Hajnoroozi, F. Aminifar, and H. Ayoubzadeh, "Generating unit model validation and calibration through synchrophasor measurements," \textit{IEEE Trans. Smart Grid}, vol. 6, no. 1, pp. 441--449, 2015.

\bibitem{Chen:2016}
Y. C. Chen, J. Wang, {A. D. Dom\'{i}nguez-Garc\'{i}a}, and P. W. Sauer, " Measurement-based estimation of the power flow Jacobian matrix," \textit{IEEE Trans. Smart Grid}, vol. 7, no. 5, pp. 2507--2515, 2016.

\bibitem{PLi18a}
P. Li, H. Su, C. Wang, Z. Liu, and J. Wu, "PMU-based estimation of voltage-to-power sensitivity for distribution networks considering the sparsity of Jacobian matrix," \textit{IEEE Access}, vol. 6, pp. 31307--31316, 2018.

\bibitem{Demarco:2016}
J. M. Lim and C. L. DeMarco, "SVD-based voltage stability assessment from phasor measurement unit data," \textit{IEEE Trans. Power Syst.}, vol. 31, no. 4, pp. 2557--2565, 2016.

\bibitem{Wangxz:2017TRWS}
X. Wang, J. Bialek, and K. Turitsyn, "PMU-based estimation of dynamic state {Jacobian} matrix and dynamic system state matrix in ambient conditions," \textit{IEEE Trans. on Power Syst.}, vol. 33, no. 1, pp. 681--690, 2018.

\bibitem{Wangxz:2017ISCAS}
X. Wang and K. Turitsyn, "PMU-based estimation of dynamic state Jacobian matrix," \textit{in Proc. IEEE Int. Symp. Circuits Syst. (ISCAS)}, Baltimore, MD, USA, May 2017, pp. 28--31.

\bibitem{Gardiner:2009}
C.  Gardiner, \textit{Stochastic Methods: A Handbook for the Natural and Social Sciences.} Berlin, Germany: Springer-Verlag Berlin Heidelberg, 2009.

\bibitem{Wangxz:CAS}
X. Wang and H. D. Chiang, "Analytical studies of quasi steady-state model in power system long-term stability analysis," \textit{IEEE Trans. Circuits and Syst. I: Regular Papers}, vol. 61, no. 3, pp. 943--956, March 2014.

\bibitem{Bialek:book}
J. Machowski, J. Bialek, and J. R. Bumby, \textit{Power system dynamics and stability.}
Chichester, U.K.: John Wiley \& Sons, 1997.

\bibitem{Pal:2010}
R. Singh, B. C. Pal, and R. A. Jabr, "Statistical representation of distribution system loads using Gaussian mixture model," \textit{IEEE Trans. Power Syst.}, vol. 25, no. 1, pp. 29--37, 2010.

\bibitem{Crow:2013}
O. A, Theresa and M. L. Crow, "An analysis of power system transient stability using stochastic energy functions," \textit{International Transactions on Electrical Energy Systems}, vol. 23, no. 2, pp. 151--165, 2013.

\bibitem{Nwankpa:2000}
H. Mohammed, and C. O. Nwankpa, "Stochastic analysis and simulation of grid-connected wind energy conversion system," \textit{IEEE Trans. Energy Conver.}, vol, 15, no. 1, pp. 85--90, 2000.

\bibitem{Elahi:1983}
H. Elahi, "Transient stability of power systems with non-linear load models using individual machine energy functions," Ph.D. dissertation, Department of Electrical Engineering, Iowa State Unviersity, Ames, Iowa, 1983.

\bibitem{GHGolub12a}
G. H. Golub and C. F. Van Loan, "Matrix Computations," The Johns Hopkins University Press, Baltimore, MD, USA: Johns Hopkins Univ. Press, 2012.

\bibitem{Chow:2011}
J. Chow, P. Quinn, L. Beard, D. Sobajic, and A. Silverstein, "Guidelines for siting phasor measurement units: Version 8 June 15." \textit{2011 North American SynchroPhasor Initiative (NASPI) Research Initiative Task Team (RITT) Report}.

\bibitem{Zhou:2011}
N. Zhou, S. Lu, R. Singh, and M. A. Elizondo, "Calibration of reduced dynamic models of power systems using phasor measurement unit (PMU) data," \textit{in Proc. North Amer. Power Symp. (NAPS)}, Boston, MA, USA, Aug. 2011, pp. 1--7.

\bibitem{Angel:2003}
A. D. Angel, M. Glavic, and L. Wehenkel, "Using artificial neural networks to estimate rotor angles and speeds from phasor measurements,"
\textit{in Proc. Intell. Syst. Appl. Power Syst. (ISAP)}, 2003, pp. 1--6.

\bibitem{Liu:2011}
J. Yan, C. C. Liu, and U. Vaidya, "PMU-based monitoring of rotor angle dynamics,"
\textit{IEEE Trans. Power Syst.}, vol. 26, no. 4, pp. 2125--2133, 2011.

\bibitem{Ross:1995}
D. F. Ross and D. Frederick. \textit{Distribution Planning and Control: Managing in the Era of Supply Chain Management.} New York, NY, US: Springer US, 2015.

\bibitem{IEEE39bus}
T. Athay, R. Podmore, and S. Virmani, "A practical method for the direct analysis of transient stability," \textit{IEEE Trans. Power App. and Syst.}, vol. PAS-98, no. 2, pp. 573--584, March 1979. 

\bibitem{IEEEStandard_amd}
\textit{IEEE Standard for Synchrophasor Measurements for Power Systems-Amendment 1: Modification of Selected Performance Requirements}.
IEEE Std C37.118.1a-2014 (Amendment to IEEE Std C37.118.1-2011), pp. 1--25, April 2014.

\bibitem{Milano:PSAT}
F. Milano, "An open source power system analysis toolbox," \textit{IEEE Trans. on Power Syst.}, vol. 20, no. 3, pp. 1199--1206, 2005.

\end{thebibliography}


\begin{IEEEbiography}[{\includegraphics[width=1in,height=1.25in,clip,keepaspectratio]{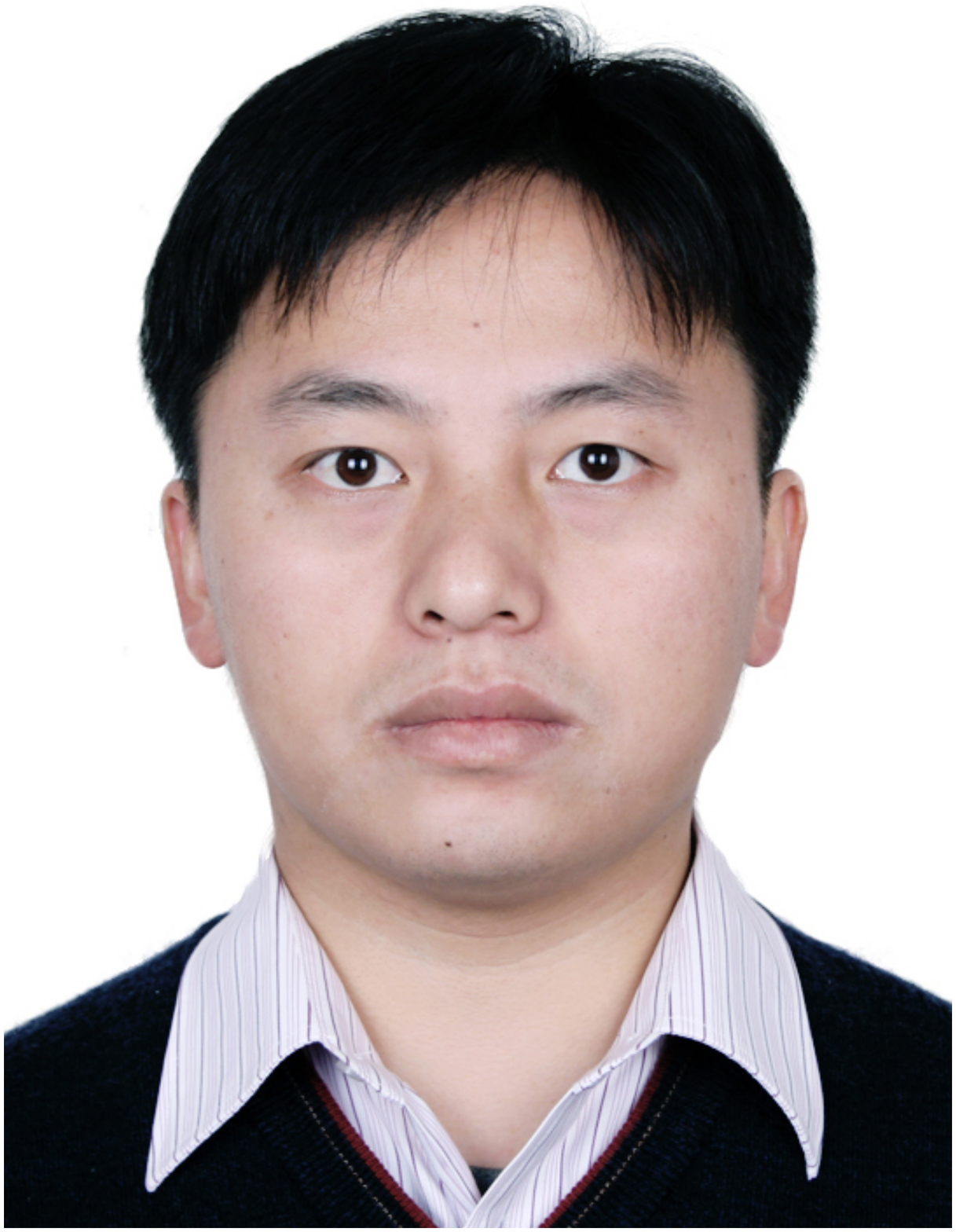}}]{Hao Sheng}
(M'14) is currently a Postdoctoral Fellow in the Department of Electrical and Computer Engineering at McGill University, Montreal, QC, Canada. He received the Ph.D. degree in the School of Electrical and Automation Engineering from Tianjin University, Tianjin, China, in 2014, the M.S. degree from Northeast Electric Power University, Jilin, China, in 2007 and the B.E. degree from North China Electric Power University, Baoding, China, in 2003, all in Electrical Engineering. From 2007 to 2012, he was affiliated with R\&D Centre of Beijing SiFang Automation Co., Ltd., Beijing, China, working on the development of PMU data-enhanced applications for Energy Management System (EMS) and Dynamic Security Assessment (DSA). From 2014 to 2017, he was a Postdoctoral Fellow in the School of Electrical and Computer Engineering at Cornell University, Ithaca, NY, USA. His research interests are in power system stability analysis and simulation, uncertainty quantification and control, and their applications in power system static and dynamic security assessment.
\end{IEEEbiography}

\begin{IEEEbiography}[{\includegraphics[width=1in,height=1.25in,clip,keepaspectratio]{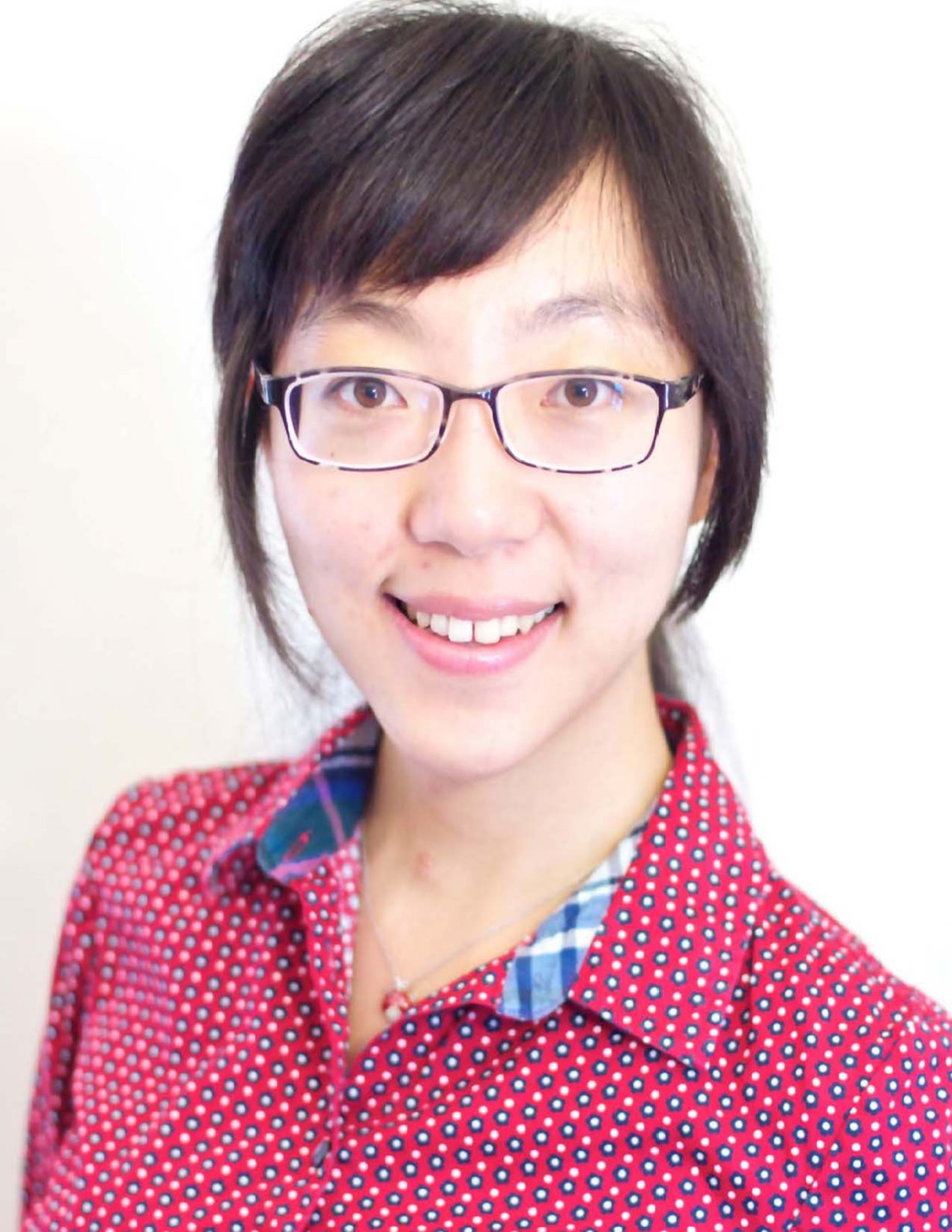}}]{Xiaozhe Wang}
is currently an Assistant Professor in the Department of Electrical and Computer Engineering at McGill University, Montreal, QC, Canada. She received the Ph.D. degree in the School of Electrical and Computer Engineering from Cornell University, Ithaca, NY, USA, in 2015, and the B.S. degree in Information Science \& Electronic Engineering from Zhejiang University, Zhejiang, China, in 2010. Her research interests are in the general areas of power system stability and control, uncertainty quantification in power system security and stability, and wide-area measurement system (WAMS)-based detection, estimation, and control.
\end{IEEEbiography}

\end{document}